\documentclass[sigconf]{acmart}
\usepackage{amsmath}
\usepackage{listings}
\usepackage{comment}
\usepackage{multirow}
\usepackage{mfirstuc}
\usepackage{textcase}
\usepackage{algorithm}
\usepackage[noend]{algpseudocode} 
\usepackage{hyperref}
\usepackage{enumerate}
\usepackage{wasysym}
\usepackage{varwidth}
\usepackage{tikz}
\usepackage{booktabs}
\usepackage{soul}
\usepackage{proof}
\usepackage{breqn}
\usepackage{makecell}
\usepackage{colortbl}
\usepackage{multicol}
\usepackage{version}

\usepackage{fancybox,graphicx}
\usepackage{adjustbox}

\excludeversion{ConferenceVersion}
\includeversion{ExtendedVersion}

\usetikzlibrary{automata,arrows,positioning,shapes.geometric,decorations.pathmorphing}
\usetikzlibrary{arrows.meta}

\makeatletter
\algrenewcommand\ALG@beginalgorithmic{\small}
\makeatother

\newcommand{\boldparagraph}[1]{\vskip 0.05in\noindent\textbf{#1.}}

\newcommand{\defi}[1]{\textit{DEF}(#1)}
\newcommand{\dep}[1]{\textit{DEP}(#1)}

\newcommand{\domain}{D}
\newcommand{\domainUnder}{\domain_{\downarrow}}

\newcommand{\groundinput}{\phi}
\newcommand{\ibs}{{\sffamily LEGOS}\xspace}

\newcommand{\approach}{{\sffamily LEGOS-PROOF}\xspace}

\newcommand{\groundAlg}{\textsc{G}}

\newcommand{\tfol}{FOL$^*$\xspace}
\newcommand{\tfolraw}{\text{FOL}^*\xspace}

\newcommand{\nf}[1]{{#1}}
\newcommand{\blue}[1]{\textcolor{blue}{#1}}

\definecolor{mygray}{rgb}{0.5,0.5,0.5}
\definecolor{mymauve}{rgb}{0.1,0.2,0.7}
\definecolor{og}{cmyk}{.6,.4,0.8,0}
\definecolor{apricot}{rgb}{0.98, 0.81, 0.69}
\definecolor{babypink}{rgb}{0.96, 0.76, 0.76}
\definecolor{beaublue}{rgb}{0.74, 0.83, 0.9}
\definecolor{flavescent}{rgb}{0.97, 0.91, 0.56}
\definecolor{lightapricot}{rgb}{0.99, 0.84, 0.69}
\definecolor{lightblue}{rgb}{0.68, 0.85, 0.9}
\definecolor{lightsalmonpink}{rgb}{1.0, 0.6, 0.6}
\definecolor{mistyrose}{rgb}{1.0, 0.89, 0.88}
\definecolor{pastelyellow}{rgb}{0.99, 0.99, 0.59}
\definecolor{mustard}{rgb}{1.0, 0.86, 0.35}
\definecolor{aureolin}{rgb}{0.99, 0.93, 0.0}
\usepackage{xspace}

\newcommand{\folstarplus}{FOL$^{*+}$\xspace}

\newcommand{\class}{\textsc{cls}}

\copyrightyear{2024}
\acmYear{2024}
\setcopyright{acmlicensed}\acmConference[ASE '24]{39th IEEE/ACM International Conference on Automated Software Engineering }{October 27-November 1, 2024}{Sacramento, CA, USA}
\acmBooktitle{39th IEEE/ACM International Conference on Automated Software Engineering (ASE '24), October 27-November 1, 2024, Sacramento, CA, USA}
\acmDOI{10.1145/3691620.3695522}
\acmISBN{979-8-4007-1248-7/24/10}

\begin{document}

\newcommand{\affilUT}{\affiliation{%
  \institution{University of Toronto}
   \city{Toronto}
    \country{Canada}
}}

\author[]{Nick Feng \hskip 0.3in Lina Marsso \hskip 0.3in Marsha Chechik}
\email{{fengnick,lmarsso,chechik}@cs.toronto.edu}
\affilUT

\title{Diagnosis via Proofs of Unsatisfiability for  First-Order Logic with Relational Objects}
%
%
\author{}


%
%


\begin{abstract}
Satisfiability-based automated reasoning is an approach that is being successfully used in software engineering to validate complex software, including for safety-critical systems.
Such reasoning underlies many validation activities, from requirements analysis to design consistency to test coverage.  
  While generally effective, the back-end constraint solvers are often complex and inevitably error-prone, which threatens the soundness of their application.  Thus, such solvers need to be validated, which includes checking correctness and explaining (un)satisfiability results returned by them. 
In this work, we consider satisfiability analysis based on First-Order Logic with relational objects (FOL*) which has been shown to be effective for reasoning about time- and data-sensitive early system designs. We tackle the challenge of validating the correctness of FOL* unsatisfiability results and deriving diagnoses to explain the causes of the unsatisfiability.
Inspired by the concept of proofs of UNSAT from SAT/SMT solvers, we define a proof format and proof rules to track the solvers' reasoning steps as sequences of derivations towards UNSAT. We also propose an algorithm to verify the correctness of FOL* proofs while filtering unnecessary derivations and develop a proof-based diagnosis to explain the cause of unsatisfiability. We implemented the proposed proof support on top of the state-of-the-art FOL* satisfiability checker to generate proofs of UNSAT and validated our approach by applying the proof-based diagnoses to explain the causes of well-formedness issues of normative requirements of software systems.

\end{abstract}

\keywords{Formal Methods for Software Engineering}

\maketitle

\section{Introduction}
\label{sec:introduction}

{Satisfiability-based automated reasoning is an approach that is being successfully used in software engineering to validate complex software, including for safety-critical systems~\cite{Matos-et-al24}.
Such reasoning underlies many validation activities, from requirements analysis~\cite{feng-et-al-23,feng-et-al-24,feng-et-al-23-b,krishna2019checking,DBLP:conf/ifm/KrishnaPS17} to design consistency~\cite{Hui-et-al-15} to test coverage~\cite{Abad-et-al-24}.}

{In a satisfiability-based approach, we  first encode a target analysis problem  as a set of constraints whose satisfiability is checked by a constraint solver. The solver then returns either a solution under which the set of constraints is satisfiable, or UNSAT, indicating that satisfiability cannot be achieved. Such solver outputs can then be used to solve the target problem. For instance, in the context of test generation via symbolic execution~\cite{King-76}, a satisfying solution to a path condition can be used to compute a concrete test input under which the path would be executed, while an UNSAT result means that the path is not feasible.} 

With the rapid advancement in the field of constraint satisfiability solving, the solvers have become progressively more advanced but also more complex and inevitably error-prone~\cite{DBLP:conf/pldi/WintererZS20}. The errors in  solvers threaten the overall soundness of the satisfiability-based reasoning approached, especially if the solver erroneously declares  UNSAT for satisfiable constraints since we cannot validate the correctness of UNSAT by checking satisfying solutions. Therefore, it is important to validate and explain  results returned by the solver. 

In this work, we consider an automated reasoning approach based on a variant of first-order logic, \tfol (first-order logic with quantifiers over relational objects)~\cite{feng-et-al-23}. \tfol has been used to specify time- and data-sensitive declarative software requirements~\cite{feng-et-al-23} including normative requirements: social, legal, ethical, empathetic, and cultural (SLEEC)~\cite{feng-et-al-23-b,feng-et-al-24,feng-et-al-24b}. An \tfol satisfiability checking approach, \ibs\cite{feng-et-al-23}, has been developed to enable automated reasoning  without the need to bound the domain for time and data, and this approach has been applied to case studies across various domains including transportation, environmental management, health, and social care~\cite{feng-et-al-23, feng-et-al-23-b, feng-et-al-24, feng-et-al-24b}.

\tfol guarantees a finite-domain property, meaning that every satisfying solution to an \tfol formula is contained within a finite domain and can thus be effectively checked for correctness. However, verifying  correctness of an \emph{unsatisfiability} result for an \tfol formula remains challenging since the result must be established across every finite but unbounded domain.  
Inspired by the concept of the proof of UNSAT from SAT/SMT solvers, in this paper, we introduce a method to efficiently generate and check the proof of unsatisfiability for \tfol. Our method defines proof rules for \tfol to capture derivation steps for constructing sound over-approximations in decidable first-order logic, and leverages the unsatisfiability of the over-approximation to prove the unsatisfiability of the original formula.

For example, consider a system that tries to match robots with humans on a field of one-dimentional coordinates. We say that a robot and a human match if the sum of their coordinates is 0. From the requirement stating that every human must be matched with a robot while every robot is to the right of the rightmost human, we can prove that the coordinate of the rightmost human is no larger than 0. We do so by proving the unsatisfiability of the following \tfol formula $\phi$, where $h_*$ and $r_*$ represent instances of humans and robots, respectively, and $t$ is the attribute storing the coordinate:
\begin{equation*}
    \begin{split}  
        \phi := \exists h_1 (\forall h_2 \cdot h_1.t 
\ge h_2.t \wedge (\exists r_1 \cdot r_1.t = -h_2.t)) \wedge \\  (\forall r_2 \cdot r_2.t > h_1.t) \wedge (h_1.t > 0)
\end{split}
    \end{equation*}
    
We can prove the unsatisfiability of $\phi$ by \textit{soundly} deriving the over-approximation clauses $\psi_6$, $\psi_7$, and $r^o_1.t > h^o_1.t$ following the proof in Fig.~\ref{fig:proof} and showing that clauses $\psi_6$, $\psi_7$, and $r^o_1.t > h^o_1.t$ together are UNSAT. The derivation of the over-approximation is recorded step-by-step to create the proof of UNSAT (see Fig.~\ref{fig:proof}), and each derivation step can be checked against \tfol derivation rules for correctness.

In addition to validating the \emph{correctness} of \tfol unsatisfiability, the proof can also be used to generate diagnoses to explain the \emph{causes} of unsatisfiability. This can be achieved by projecting the derivation steps in the proof back onto the input formula. The projection result highlights the atoms that are active for the derivation of UNSAT and enables us to slice away inactive parts of the formula without affecting the overall unsatisfiability. In the above example, projecting the UNSAT proof of $\phi$ shown in Fig.~\ref{fig:proof}
indicates that the atom $\forall h_2 \cdot h_1.t > h_2.t$ is inactive, and hence can be sliced away in the diagnosis without affecting the unsatisfiability of $\phi$. Therefore, the diagnosis of the unsatisfiability of $\phi$ is
\begin{equation*}
    \phi_{dig} := \exists h_1 (\forall h_2 (\exists r_1 \cdot r_1.t = -h_2.t)) \wedge  (\forall r_2 \cdot r_2.t > h_1.t) \wedge (h_1.t  > 0)
\end{equation*}
This means we can prove that every human (instead of just the rightmost human) must be at a coordinate ($h_1.t$) greater than or equal zero.
\boldparagraph{Contributions}
We make the following theoretical contributions:  
a proof format for \tfol unsatisfiability; a proof checking algorithm to verify the correctness of \tfol proof while slicing away unnecessary derivations; and an application of \tfol proofs to generate proof-based diagnoses to explain  causes of unsatisfiability. 
We also make the following engineering contributions:  an extension of
the state-of-the-art FOL* satisfiability checker to generate \tfol proofs of UNSAT and an application of proof-based diagnoses to a software engineering problem of addressing causes of well-formedness violations of normative requirements.  
Specifically, we use eight case studies to show that proofs of unsatisfiability can lead to effective debugging of conflicting, redundant, and overly restrictive normative requirements.

\boldparagraph{{Significance}}
{This paper proposes a methodology to generate and verify proofs of unsatisfiability for \tfol. The proposed methodology is important for explaining the cause of unsatisfiability to both technical and non-technical stakeholders, enabling them to diagnose and resolve requirements-level issues. Our approach is not limited to this application domain; it also serves as a bridge to assist other areas where \tfol analysis is employed, further supporting users without formal methods expertise.
}

The rest of the paper is organized as follows: 
Sec.~\ref{sec:background} gives the background material for our work.
Sec.~\ref{sec:proof} introduces the proof framework, \approach, to generate proofs of unsatisfiability for \tfol.
Sec.~\ref{sec:proofchecking} describes a method to check correctness of such proofs.  
Sec.~\ref{sec:dignoise} describes how to use such proofs to compute diagnoses of \tfol unsatisfiability.  
Sec.~\ref{sec:LEGOsSupport} presents the integration of our proof support with the state of the art \tfol solver.
Sec.~\ref{sec:evaluation} presents the evaluation of the approach's efficiency and effectiveness.
Sec.~\ref{sec:relatedwork} compares \approach{} to related work.
Sec.~\ref{sec:conclusion} summarizes the paper and discusses future research.

%
\section{Background}
\label{sec:background}
In this section, we briefly describe  \tfol~\cite{feng-et-al-23} and its satisfiability. 

We start by introducing the syntax of \tfol. 
A \emph{signature} $S$ is a tuple $(C, R, \iota)$, where $C$ is a set of constants, $R$ is a set of relation symbols, and $\iota : R \rightarrow \mathbb{N} $ is a function that maps a relation to its arity.  We assume that the domain of constant $C$ is $\mathbb{Z}$, where the theory of linear integer arithmetic (LIA) holds. Let $V$ be a 
set of variables in the domain $\mathbb{Z}$. A \emph{relational object} $o:r$ of class $r \in R$ is an object with $\iota(r)$ regular attributes and one special attribute \textit{ext}, where every attribute is a variable. We assume that all regular attributes are ordered and let $o[i]$ denote the $i$th attribute of $o$. 
Some attributes are named, and $o.x$ refers to $o$'s attribute with the name `$x$'.  
Each relational object $o$ has a special attributes: $o.ext$ indicating whether $o$ exists in a solution. For convenience, we define a function $\class$($o:r$) to return the relational object's class $r$.  Let \emph{a term} $t$ be defined inductively as $t: \; c \; | \;  var \; | \; o[i] \; | \; o.x \; | \; t + t \; | \;  c \times t$ for any constant $c \in C$, any variable $var \in V$, any relational object $o$, any index $i \in [0, \iota(r)]$ and any valid attribute name $x$. Given a signature $S$, the syntax of the \tfol{} formulas is defined as follows:
    \emph{(1)} $\top$ and $\bot$, representing values ``true'' and ``false''; 
    \emph{(2)} $t = t'$ and $t > t'$, for term $t$ and $t'$;  
    \emph{(3)} $\phi \wedge \psi$, $\neg \phi$ for \tfol{} formulas $\phi$ and $\psi$; 
    \emph{(4)} $\exists o:r \cdot  (\phi)$ for an \tfol{} formula $\phi$  and a class $r$;
    The quantifiers for \tfol{} formulas are limited to relational objects, as shown by rules (4).  
    Operators $\vee$ and $\forall$ can be defined in \tfol as follows: $\phi \vee \psi = \neg (\neg \phi \wedge \neg \psi)$ and $\forall o:r \cdot \phi= \neg \exists o:r \cdot \neg \phi$. We say an \tfol formula is in a \textit{negation normal form} (NNF) if negations ($\neg$) do not appear in front of $\neg$, $\wedge$, $\vee$, $\exists$ and $\forall$. For the rest of the paper, we assume that every \tfol $\phi$ formula is in NNF.

Let $\domain$ be a domain of relational objects and $\phi$ be an \tfol formula. A valuation function $v$ assigns concrete values to (1) every free variable in $\phi$ and (2) every attribute variable for every relational object in $\domain$, including Boolean attribute $ext$. Let $v(var)$ denote the value assigned to a variable $var$.  With slight abuse of notation, we define $v(t, \domain)$ to be the result of evaluating  a term $t$ in a domain $\domain$ with the following rules:

\begin{align*}
    v(t, \domain) = \begin{cases}
  c  & \text{if } t \text{ is a constant $c$} \\
  v(var) & \text{if } t \text{ is a variable $var$}  \\
  -v(t', \domain) & \text{if } t = -t' \\
  v(t_1, \domain) \; op \; v(t_2, \domain) & \text{if } t = t_1 \; op \; t_2 \text{ for } op \in \{+ , \times\}
\end{cases}
\end{align*}

\begin{figure}[t]
\begin{tabular}{lcl}
            \centering
            $(\domain, v) \models \top$ & & $(\domain, v) \not\models \bot$ \\
            $(\domain, v) \models o.ext$ & iff & $v(o.ext) = \top$ \\
              $(\domain, v) \models t = t'$ & iff & $v(t, \domain) = v(t', \domain)$\\
              $(\domain, v) \models t > t'$ & iff & $v(t, \domain) > v(t', \domain)$\\
             $(\domain, v) \models \neg \phi$ & iff & $(\domain, v) \not\models \phi$ \\
             $ (\domain, v) \models \phi \wedge \psi $ & iff & $(\domain, v) \models \phi$ and  $(\domain, v) \models \psi$\\
             $ (\domain, v) \models \exists o:r \cdot \psi $ & iff & $ (\domain, v) \models \bigvee_{o' \in \domain_r} (o'.ext \wedge \psi[o \gets o'])$ 
\end{tabular}
\vspace{-0.1in}
\caption{{\small \folstarplus semantics. $\domain_r$ defines the set of  all relational objects of class $r$ in $\domain$.}}
\label{tab:semantic}
\vspace{-0.2in}
\end{figure}

Given an \tfol formula $\phi$, $(\domain, v) \models \phi$ means that $\phi$ holds in a domain $\domain$ with respect to a valuation function $v$. The semantics of $(\domain, v) \models \phi$ is defined in Fig.~\ref{tab:semantic}. The semantic rule for quantified formula (i.e., $\exists o: r \cdot \: (\phi)$) expands the quantifier on relational objects according to the domain $\domain$ and yields the expression $\bigvee_{o':r}^{\domain} (o'.ext \wedge \phi [o \leftarrow o'])$. By recursively expanding the \tfol formula (and its sub-formulas) in a given domain $\domain$, we obtain a quantifier-free formula with a bounded number of free variables, referred to as the \textit{grounded} formula in $\domain$. The semantic relation ($\domain$, $v$) $\models \phi$ holds  in the \tfol formula if the grounded formula of $\phi$ in $\domain$ evaluates to $\top$ according to $v$. {A $\phi$ formula is satisfiable if there \textit{exists} a tuple ($\domain$, $v$) such that ($\domain$, $v$) $\models \phi$.  ($\domain$, $v$) is referred as \emph{a solution} to $\phi$. }
Given a solution $\sigma = (D, v)$, we say that a relational object $o$ is in $\sigma$, denoted $o \in \sigma$, if $o \in D$ and $v(o.ext)$ is true. {Moreover, two relational objects $o_1$ and $o_2$ are equivalent ($o_1 \equiv o_2$) in a solution if $o_1$ and $o_2$ are object of the same class $r$, and for every attribute $x$ of class $r$, $v(o_1.x) = v(o_2.x)$}. The \emph{volume of the solution}, denoted as $vol(\sigma)$, {is the size of the set of unique relational objects in the solution: $|\{o \mid o \in \sigma\}|$ where $\forall o_1, o_2 \in \sigma: o_1 \not\equiv o_2$.}
\begin{example}
Let $a$ be a relational object of class $\textit{A}$ with an attribute name $val$. The formula $\forall a: A.\: (\exists a':A \cdot \: (a.val < a'.val) \wedge \exists a'':A \cdot \: a''.val = 0) $ has no satisfying solutions in any finite domain. On the other hand, the formula $\forall a:A \cdot \: (\exists a', a'':A  \cdot \: (a.val = a'.val + a''.val) \wedge \exists a^*:A \cdot \: a^*.val = 5) $ has a solution  $\sigma = (\domain, v)$ of volume 2, with the domain $\domain = (a_1, a_2)$ and the value function $v(a_1.val = 5)$, $v(a_2.val = 0)$ because if $a \gets a_1$ then the formula is satisfied by assigning $a' \gets a_1, \;a'' \gets a_2, \; a^* \gets a_1$; and  if $a \gets a_2$, then the formula is satisfied by assigning $a' \gets a_2, \;a'' \gets a_2,\; a^* \gets a_1$.
\end{example}

\section{Proof of UNSAT}\label{sec:proof}
Here we first introduce the proof framework (Sec.~\ref{sec:proofFramework})  that explains the syntax and semantics of a \tfol proof. Then we will explain how the proof framework interact with external theory reasoning via interface derivation rules in Sec.~\ref{ssec:interface}. Finally we instantiate the framework with derivation rules specific to \tfol  (Sec.~\ref{sec:folrules}). 

\subsection{\tfol Proof of Unsatisfiability}\label{sec:proofFramework}

In this section, we present the framework for \tfol proof which is parameterized on the use of an external theory $\mathbb{T}$. 

Let
$\mathbb{T}$ be a quantifier-free external theory (or a theory combination), e.g., quantifier-free linear integer arithmetic (QFLIA), for which a decision procedure exists. We assume that $\mathbb{T} \subseteq \tfolraw$ and  $\mathbb{T} \cap \tfolraw$ on \textit{at least} boolean logic symbols (i.e., $\top$, $\bot$, $\wedge$, $\neg$) with consistent interpretations.  The syntax and semantics of \tfol presented in Sec.~\ref{sec:background} assume that
$\mathbb{T}$ is QFLIA. However,  \tfol can be extended with other background theories such as quantifier-free linear arithmetic on reals (QFLRA), mixed linear arithmetic (QFLIRA) {or any quantifier-free background theory that satisfies the above requirements.}

Let $R$ be a set of input \tfol constraints. A \emph{\textbf{UNSAT} proof of $R$} is a sequence of derivation steps $d_1, \ldots, d_n$ where every step $d_i$ ({formally defined later}) is an application of one of the derivation rules that transforms the proof state $s_{i-1}$ into a new state $s_i$. A \emph{proof state} $s_i$ is a tuple $(R^+_i, F_i, \domain_i)$, where
$R^+_i$ is a set of formulas in \tfol ("lemmas");
$F_i$ is the set of formulas in
$\mathbb{T}$ ("facts"); and $D_i$ is a set of side-effect objects (SEOs) used to 
capture the side-effects of the derivations (e.g., \tfol relational objects). Initially,  $R^+_0 = R$ and $F_0 = D_0 = \emptyset$.

Each \emph{derivation step} $d_i$  is a tuple $(\textit{rule}, Dep_i, F_i^\Delta, R_i^\Delta, D_i^\Delta)$, where \textit{rule} is the name of the derivation rule being applied;  $Dep_i = (R', F', D')$ is  sets of dependant lemmas $R'_i$, facts $F'_i$ and SEOs $\domain'_i$ that enable the derivation; $F_i^\Delta$ is a set of new facts added by applying $d_i$; $R_i^\Delta$ is a set of new lemmas derived by $d_i$, and $D_i^\Delta$ are new SEOs added by $d_i$. We say that dependencies  $Dep_i$ are \textit{satisfied} in state  $s_{i-1}$, denoted as $Dep_i \subseteq s_{i-1}$, if $R'_i \subseteq R^+_i$, $F'_i \subseteq F_{i-1}$, and $\domain'_i \subseteq \domain_{i-1}$.
Let a state $s_{i-1} = (F_{i-1}, R^+_{i-1}, D_{i-1})$ be given. If $Dep_i \subseteq s_{i-1}$ 
 and the \textit{rule}-specific enabling condition, denoted as $PC_{\textit{rule_i}}$, holds on $d_i$, then applying $d_i$ adds the newly derived resources to $ F_i^\Delta$, $R_i^\Delta$, and $D_i^\Delta$ to $F_{i-1}$, $R^+_{i-1}$, and $D_{i-1}$, respectively.
Formally, the application of $d_i = (\textit{rule}_i, Dep_i, F_i^\Delta, R_i^\Delta, D_i^\Delta)$ in the state $s_{i-1} = (F_{i-1}, R^+_{i-1}, D_{i-1})$, denoted as $s_{i-1}\circ d_i$, yields a new state $s_i = (F_{i}, R^+_{i}, D_{i}) = $ 
\[
\begin{cases}
 \emptyset & \text{if } s_{i-1} = \emptyset \\
  (F_{i-1} \cup F_i^\Delta, R^+_{i-1} \cup R_i^\Delta, D_{i-1} \cup D_i^\Delta )  & \text{if } Dep_i \subseteq s_{i-1} \wedge \textit{PC}_{rule_i}(d_i) \\
  \emptyset & \text{otherwise}
\end{cases}
\]
where $\emptyset$ is a dead-end state representing a derivation error, and function \textit{PC}$(rule_i, Dep_i)$ checks sufficiency of the listed resources in $Dep_i$ to satisfy the enabling condition specific to each \textit{rule_i}.

\begin{example}
    {Suppose we introduce a derivation rule {\textbf{UP}}  
 allowing to derive  new facts via unit propagation in propositional logic. More specifically, given a derivation step $d_i:= \{{\textbf{UP}}, R^{\Delta}_i, F^{\Delta}_i, \domain^{\Delta}_i\, Dep' := \{R'_i, F'_i, \domain'_i\})$, the derivation condition is $PC_{\textbf{UnitProp}}(d_i) = F'_{i-1} \vdash_1 F^{\Delta}_i \wedge R^{\Delta}_i = \domain^{\Delta_i} = \emptyset$, 
    where $F'_{i-1}  \vdash_1 F^{\Delta}_i$ means that every fact in $F^{\Delta}_i$ can be derived from facts in $F'_{i-1}$ by unit propagation.  Let a proof state $s_{i-1} := \{R_{i-1}, F_{i-1}, \domain_{i-1}\}$  where $\{a, \neg a \vee b, \neg b \vee c, c \vee f\} \subseteq F_i$ be given.  Then the derivation step {$d_i:= \{\textbf{UP}, F^{\Delta}_i := \{c\}, \ldots $, $Dep' := \{F'_i := \{a, \neg a \vee b, \neg b \vee c\}, \ldots \}\}$}  on $s_{i-1}$ is \emph{valid}  because (1) $Dep' \subseteq s_{i-1}$ and (2) $PC_{\textbf{{UP}}}(d_i)$ evaluates to true.  Applying $d_i$ yields the state $s_{i-1} \circ d_{i} = \{R_{i-1}, F_{i-1} \cup \{c\}, \domain_{i-1}\}$. 
    }
\end{example}

\begin{definition}[State Satisfiability]
    {Let a state $s_{i} = (F_{i}, R^+_{i}, D_{i})$ be given. The state $s_i$ is \emph{satisfiable}, denoted as $SAT(s_i)$, if the \tfol formula $(\bigwedge R^+_{i} ) \wedge (\bigwedge F_i)$ is satisfiable. An empty state $\emptyset$ is trivially satisfiable.} 
\end{definition}

\begin{definition}[Derivation Soundness]
    {Let a proof state $s_{i-1}$ and a derivation $d_i$ be given. The derivation step $d_i$ is \emph{sound} if the derived new state $s_{i-1}\circ d_i$ preserves the satisfibility of $s_i$: $SAT(s_i) \iff SAT(s_{i-1}\circ d_i)$. }
\end{definition}

\begin{theorem}[Partial Soundness]
    Let a set of \tfol formulas $R$ and a proof $L=d_1 \ldots d_n$ be given. Suppose $s_0$ is the initial proof state for $R$ and $s_n = s_0\circ d_1 \circ \ldots d_n$ is the final proof state after applying every derivation rule in $L$. If (1) every derivation step is sound, (2) $s_n:=(R_n, F_n, \domain_n) \neq \emptyset$, and (3) $\bot \in R_n $ or $\bot \in F_n$, then the conjunct of $R$ (i.e., $\wedge R$) is UNSAT. 
\end{theorem}

\subsection{Interfacing with the Background Theories}\label{ssec:interface}
Our proof framework supports derivation of unsatisfiability via external reasoning in some parameterized background theory $\mathbb{T}$. More specifically, in a proof state $s_i = (R^+_i, F_i, \domain_i)$, the set of facts $F$ is maintained separately from \tfol lemmas $R^+$ so that they can be used exclusively for the purpose of external theory reasoning. To enable such support, we need an interface to capture the derivation in $\mathbb{T}$ and to communicate between \tfol and the background theory. 

We add the following three meta derivation rules: (1) the rule \textbf{$\mathbb{T}$-Derive} which enables external reasoning in $\mathbb{T}$ based on the facts in $F$; (2) the rule \textbf{$\tfolraw \rightarrow \mathbb{T}$} which {communicates} \tfol lemmas from $R^+$ to $F$; and (3) \textbf{$\mathbb{T} \rightarrow \tfolraw$} which communicates facts from $F$ back to $R^+$. We formalize this below.
    \begin{definition}\label{rule:derive}
    Let a derivation step $d_i = (\textit{rule}_i, Dep_i, F_i^\Delta, R_i^\Delta, D_i^\Delta)$ and a formula $\phi_g$ in theory $\mathbb{T}$, where $Dep_i = (R'_i, F'_i, \domain'_i)$, be given. The step $d_i$ is a \emph{valid application of} the \textbf{$\mathbb{T}$-Derive} rule if
        $PC_{\textbf{$\mathbb{T}$-Derive}}(d_i) := R^\Delta_i = \domain^\Delta_i = \emptyset \wedge F^\Delta_i = \{\phi_g\} \wedge F'_i \models^{\mathbb{T}} \phi_g$.
    \end{definition}
{That is, if there exists a subset of facts $F' \subseteq F$ such that $F'$ derives a new fact $\phi_g$ in $\mathbb{T}$ (i.e., $F' \models^{\mathbb{T}} \phi_g$), then applying rule \textbf{$\mathbb{T}$-Derive} adds $\phi_g$ to $F_i$. The fact $\phi_g$ depends on $F'$, denoted as  $\dep{\phi_{g}} = F'$.}    
     \begin{definition}\label{rule:LtoT}
    Let a derivation step $d_i = (\textit{rule}_i, Dep_i, F_i^\Delta, R_i^\Delta, D_i^\Delta)$ and an \tfol formula $\phi$ where $ Dep_i := (R'_i, F'_i, \domain'_i)$ be given. The step $d_i$ is \emph{a valid application of the \textbf{$\tfolraw \rightarrow \mathbb{T}$} rule} if
         $PC_{\textbf{$\tfolraw \rightarrow \mathbb{T}$}}(d_i) := R^\Delta_i = \domain^\Delta_i = \emptyset \wedge F^\Delta_i = \{\phi\} \wedge \phi \in R'_i \wedge \phi \in \mathbb{T}$.
\end{definition}    
That is, if $\phi \in R^+$ is in the intersection between \tfol and $\mathbb{T}$, then $\phi$ can be added to $F$ as a $\mathbb{T}$ fact.
    \begin{definition}\label{rule:TtoL}
    Let a derivation step $d_i = (\textit{rule}_i, Dep_i, F_i^\Delta, R_i^\Delta, D_i^\Delta)$ and a $\mathbb{T}$ formula $\phi$ where $ Dep_i := (R'_i, F'_i, \domain'_i)$ be given. The step $d_i$ is \emph{a valid application of the \textbf{$\mathbb{T}  \rightarrow \tfolraw$} rule} if
         $PC_{\textbf{$\mathbb{T} \rightarrow \tfolraw$}}(d_i) := F^\Delta_i = \domain^\Delta_i = \emptyset \wedge R^\Delta_i = \{\phi\} \wedge \phi \in F'_i \wedge \phi \in \tfolraw$.        \end{definition}
{In our proof framework, we can derive \tfol lemmas with \tfol derivation rules (to be presented in Sec.~\ref{sec:folrules}). For lemmas  in the intersection between \tfol and the background theory $\mathbb{T}$, we can apply the derivation rules \textbf{$\tfolraw \rightarrow \mathbb{T}$} to add them as facts in $F$. The facts in $F$ can then be used for external reasoning via the step \textbf{$\mathbb{T}-{\bf Derive}$} which derives additional facts in $F$. The newly-derived facts can  be added as \tfol lemmas in $R^+$ via  \textbf{$\mathbb{T}\rightarrow  \tfolraw$} to make further progress for the derivation of UNSAT.
{Note that if \textbf{UNSAT} is derived via the external reasoning (\textbf{$\mathbb{T}-{\bf Derive}$}),   the \tfol proof of unsatisfiability concludes as well. In such a case, the facts in $F$ represent an over-approximation of the input \tfol formulas $R$ as they are soundly derived from $R$. The unsatisfiability of the over-approximation implies the unsatisfiability of $R$. }

\subsection{\tfol Derivation Rules}\label{sec:folrules}
In this section, we introduce \tfol-specific derivation rules which allow us to show soundness of derived \tfol lemmas.  
\begin{definition}[Definition Variable]
    Let $\phi$ be an \tfol formula. A \emph{definition variable}, denoted as $\defi{\phi}$, is a boolean variable subject to the following two constraints $\phi_{\text{def}}^+$ and $\phi_{\text{def}}^-$ where
    $\phi_{def}^+:= l \Rightarrow \phi$ and $\phi_{def}^-:= \bar{l} \Rightarrow \neg \phi$.
\end{definition}

The definition variables serve as boolean abstractions to \tfol formulas. The relationship between a definition variable $l$ and a formula $\phi$ is captured by the positive and negative \textit{definition clauses}, $\phi^+{\textit{def}}$ and $\phi^-_{\textit{def}}$, respectively. We assume that $\phi^+_{\textit{def}}$ and $\phi^-{\textit{def}}$ can be accessed via the functions $DL^+$ and $DL^-$, respectively (i.e., $DL^+(\phi) = \phi^+_{\textit{def}}$ and $DL^-(\phi) = \phi^-_{\textit{def}}$).

In a proof state $s_i = (R^+_i, F_i, \domain_i)$, the abstraction function $\textit{DEF}$ is stored as tuples in the SEO domain $D_i$. Given a domain of SEOs $\domain_i$, we say the \emph{$\phi$ is defined in $\textit{DEF}$}, denoted as $\phi \in \textit{DEF}$, if there exists a tuple $(\phi, l) \in \domain_i$. We ensure that the function $\textit{DEF}$ is one-to-one, such that the inverse function $\textit{DEF}^{-1}$ can be defined (i.e., $\phi = \textit{DEF}^{-1}(\defi{\phi})$).  In addition to the abstraction function $\textit{DEF}$, the \emph{side-effect domain} $\domain_i$ also contains relational objects whose classes are defined in the signature of the input \tfol formulas $R$. 

{The full set of \tfol-specific derivation rules is presented in Fig.~\ref{fig:folrules}, organized as the name (Name), the enabling condition ($PC_{\textbf{Name}}$), and the dependencies of each derivation rule.}
In the rest of the section, we explain some of these derivation rules. Their formal semantics is given in 
\begin{ConferenceVersion}
    the extended version~\cite{extended~ASE}.
\end{ConferenceVersion}
\begin{ExtendedVersion}
    Appendix~\ref{app:folproofsem}.
\end{ExtendedVersion}

\boldparagraph{Define} If $\phi$ is an \tfol formula and \defi{$\phi$} is undefined, then applying the \textbf{Define} rule assigns $\defi{\phi}$ to a fresh variable and adds the definition clauses $DL^+(\phi)$ and $DL^-(\phi)$ to $R^+$.
    The derivation rule \textbf{Define} allows the introduction of new \tfol formulas in the proof of UNSAT. For example, \textbf{Define} can introduce $\phi$ as a hypothesis, and by deriving $DEF(\phi)$ as a lemma (via $DL^-(\phi)$), we can then use $\phi$ as a lemma (via $DL^+(\phi)$) in the rest of the proof.

\begin{figure*}
    \centering

    \begin{tabular}{| c |c | c | }
        \toprule
        \textbf{Name} & \makecell{$PC_{\textit{Name}}(d_i)$, where $d_i = (\textit{Name}, Dep_i, F_i^\Delta, R_i^\Delta, D_i^\Delta)$ 
        and $Dep_i = \{R'_i, F'_i, \domain'_i\}$} & Dependencies\\
        \hline\hline
         \textbf{Define} & \makecell{$ R^\Delta_i = \{DL^+(\phi),  DL^-(\phi)\} \wedge F^\Delta_i = \emptyset \wedge $ $\domain^\Delta_i = \{\defi{\phi} = l\}$  and $l$ is fresh} & \makecell{$\emptyset$}
         \\ \hline
         \textbf{Subs} & \makecell{$F^\Delta_i = \domain^\Delta_i = \emptyset \wedge R^+_{i} = \{\phi^+_{subs}, \phi^-_{subs}\}$  $\wedge FV(\phi) \supseteq FV(\psi) \wedge $  $\defi{\phi} \in \domain'_i \wedge \defi{\psi}\in \domain'_i$\\
         where $\phi$ and $\psi$ are \tfol formula \\
         $\phi^+_{subs}: \defi{\phi} \Rightarrow \phi[\psi \gets \defi{\psi}]$ and 
         $\phi^-_{subs}: \overline{\defi{\phi}} \Rightarrow \neg \phi[\psi \gets \defi{\psi}]$} 
         &\makecell{$\textbf{Dep}(\phi^+_{subs}) :=$  $\{DL^+(\phi), DL^-{\psi}\}$ \\
          $\textbf{Dep}(\phi^-_{subs}) :=$ $\{DL^-(\phi), DL^+(\psi)\}$
         } \\ 
         \hline
         \textbf{ApplyLemma} & \makecell{$F^\Delta_i = \domain^\Delta_i = \emptyset \wedge R^\Delta_i = \{\defi{\phi}\} \wedge$ 
         $\defi{\phi} \in \domain'_i \wedge \phi \in R'_i$}&  \makecell{\textbf{Dep}($\defi{\phi}$) :=  
         $\phi$} \\
         \hline
         \textbf{RewriteOr} & \makecell{$F^\Delta_i = \domain^\Delta_i = \emptyset \wedge R^\Delta_i = \{\phi_{or+}, \phi_{or-}^l, \phi_{or-}^r\} \wedge $ 
         $\defi{\phi} \in \domain'_i \wedge \defi{A} \in \domain'_i \wedge \defi{B} \in \domain'_i$ \\
         where $\phi = A \vee B$ and $A$, $B$ are \tfol formula \\ 
         $\phi_{or+} := \defi{\phi} \Rightarrow \defi{A} \vee \defi{B}$\\
         $\phi_{or-}^l := \overline{\defi{\phi}} \Rightarrow \overline{\defi{A}}$ and 
         $\phi_{or-}^r := \overline{\defi{\phi}} \Rightarrow \overline{\defi{B}}$}& \makecell{$\textbf{Dep}(\phi_{or+}):=$\\
         $\{DL^+(\phi), DL^-(A), $ $DL^-(B)\}$ \\
         $\textbf{Dep}(\phi_{or-}^l):=$ $\{
         DL^-(\phi), DL^+(A)\}$\\
         $\textbf{Dep}(\phi_{or-}^r):=$ $\{
         DL^-(\phi), DL^+(B)\}$} \\
         \hline
        \textbf{ExistentialInst} & \makecell{$F^\Delta_i  = \emptyset \wedge \domain^\Delta_i =\{o':r\} \wedge R^\Delta_i = \{\phi_{o'}^+, \phi_{o'}^-\} \wedge$ 
        $\phi = \exists o:r\cdot p(o)$ and $\defi{\phi} \in \domain'_i$ \\ 
        where $o'$ is a fresh relational object of class $r$ \\
        $\phi_{o'}^+ = \defi{\phi} \Rightarrow (o'.ext \wedge p(o'))$ and  
        $\phi_{o'}^- = \overline{\defi{\phi}} \Rightarrow \neg (o'.ext \wedge p(o'))$} & \makecell{$\textbf{Dep}(o'):= \{DL^+(\phi)\}$\\
        $\textbf{Dep}(\phi_{o'}^+):= \{DL^+(\phi)\}$
        \\
       $\textbf{Dep}(\phi_{o'}^-):= \{DL^-(\phi)\}$ }\\
        \hline
         \textbf{UniversalInst} & \makecell{$F^\Delta_i = \domain^\Delta_i = \emptyset \wedge R^\Delta_i = \{\phi_{o'}\} \wedge $
         $\phi = \forall o:r \cdot p(o) \wedge \defi{\phi} \in \domain'_i \wedge o':r \in \domain'_i$ \\ 
         where $\phi_{o'} = \defi{\phi} \Rightarrow (o'.ext \Rightarrow p(o'))$} & \makecell{$\textbf{Dep}(\phi_{o'}) :=$ 
         $\{o', DL^+(\phi)\}$} \\
         \hline
         \textbf{RewriteNeg} & \makecell{$F^\Delta_i = \domain^\Delta_i = \emptyset \wedge R^\Delta_i = \{ \phi_{neg}^+, \phi_{neg}^-\} \wedge$ 
         $\defi{\phi} \in \domain'_i$ \\ where $\phi = \neg \psi$, $\phi_{neg}^+ := \defi{\phi} \Rightarrow NEG(\psi)$ and 
         $\phi_{neg}^- := \overline{\defi{\phi} }\Rightarrow \psi$ \\
         where $\textit{NEG}(\neg \psi) = \psi, \; \textit{NEG}(\psi_1 \wedge \psi_2) = \neg \psi_1 \vee \neg \psi_2, \; \textit{NEG}(\psi_1 \vee \psi_2) = \neg \psi_1 \wedge \neg \psi_2$ \\
         $\textit{NEG}(\exists o:r\cdot P(o)) = \forall o:r\cdot \neg P(o),  \; \textit{NEG}(\forall o:r\cdot P(o)) = \exists o:r\cdot \neg P(o), \; \textit{NEG}(\psi) = \neg \psi
    $
         } & \makecell{$\textbf{Dep}(\phi_{neg}^+):=$
         $\{DL^+(\phi)\}$ \\
         $\textbf{Dep}(\phi_{neg}^-):=$
         $\{DL^-(\phi)\}$ }\\
         \hline
         \textbf{Unit} & \makecell{$F^\Delta_i = \domain^\Delta_i = \emptyset \wedge R^\Delta_i = \{ \psi \} \wedge$ 
         $\phi = l \Rightarrow \psi \wedge \phi \in R'_i \wedge l \in R'_i$ }& $\textbf{Dep}(\psi) = \{\phi, l\}$\\
         \hline 
         \textbf{UNSAT}& \makecell{$\bot \in R'_i \vee \bot \in F'_i$} & 
         \makecell{$\bot$}\\
         \hline
    \end{tabular}
    
    \caption{{\small The full list of \tfol-specific derivation rules with their names, enabling condition $PC_{\textit{Name}}$, and the dependencies of the derived resources (i.e., $\textbf{Dep}(r)$ where $r$ is a resource derived by an application of the rule  \textit{Name}).   $\overline{l}$ refers to the complement of the literal $l$ (e.g., $\overline{\defi{\phi}}$). 
    } 
    } 
    \label{fig:folrules}
    \vspace{-0.15in}
\end{figure*}

\boldparagraph{RewriteNeg} Suppose an \tfol formula matches the pattern $\neg \psi$, where $\psi$ is also an \tfol formula. If $\defi{\phi}$ is defined, then applying \textbf{RewriteNeg} on $\phi$ adds the following lemmas to $R^+$: (1)
    $\phi_{neg}^+ := \defi{\psi} \Rightarrow \textit{NEG}(\phi)$ and (2)
     $\phi_{neg}^- := \overline{\defi{\phi}} \Rightarrow \psi$,  
    where the function \textit{NEG}  pushes the top-level negation into $\psi$ and is defined in a standard way as shown in Fig.~\ref{fig:folrules}.

{Note that applying \textbf{RewriteNeg} recursively converts an \tfol formula into its negation normal form (NNF) where negation symbols appear in front of atoms (e.g., terms or literals). We can also express \textbf{RewriteAnd} with \textbf{RewriteOr} and \textbf{RewriteNeg}.}

    \boldparagraph{UNSAT} If $\bot$ has been derived as a fact or lemma in the current state, then the {\bf UNSAT} rule can signal the end of the proof.

Each derivation rule simulates a key reasoning step of \tfol satisfiability: the \textbf{Define} and \textbf{Substitute} rules abstract complex \tfol formulas and capture their propositional relations.  The \textbf{RewriteOr} and \textbf{RewriteAnd} rules handle case splitting. The \textbf{UniversalInst} and \textbf{ExistentialInst} rules handle quantifiers over relational objects, and \textbf{RewriteNegn} handles negation and enables NNF conversion. The \textbf{ApplyLemma} rule enforces the asserted true statements, and \textbf{Unit} enables lemma simplifications through resolution.

\subsection{Derivation Macros}\label{sec:shortcut}
{The \tfol proof rules are fine-grained and  sometimes cumbersome to write}. We noticed that there are common rule usage patterns  that allow us to {chain} \tfol derivations to form macros to make the proof succinct:

\boldparagraph{RewriteAND*} If a formula $\phi \in R^+$ matches $\psi_1 \wedge \psi_2  \ldots \wedge \psi_n$, then  $\psi_1$, $\psi_2 \ldots \psi_n$ can be directly added to $R^+$. The expansion of the macro \textbf{RewriteAND*} is 
\begin{ExtendedVersion}
    shown in Appendix~\ref{ap:marcoexpansion}.
\end{ExtendedVersion}
\begin{ConferenceVersion}
    shown in the extended version~\cite{extended_ASE}.
\end{ConferenceVersion}

    \boldparagraph{RewriteOR*} If a formula $\phi \in R^+$ matches $\psi_1 \vee \psi_2 \ldots \vee $, then the positive definition lemma $DL^+(\phi_1) \ldots L^+(\phi_n) $ and the lemma
    $\defi{\psi_1} \vee \defi{\psi_2} \ldots \vee \defi{\psi_n}$  can be directly added to $R^+$. 
    
    \boldparagraph{UniversalInst*} If a formula $\phi \in R^+$ matches $\forall o:r p(o)$ and there exists a relational object $o'$ of class $r$, then $o'.ext \Rightarrow p(o')$ can be added to $R^+$.
    
    \boldparagraph{ExistenialInst*} If a formula $\phi \in R^+$ matches $\exists o:r p(o)$, then add a fresh relational object $o'$ to $\domain$ and a lemma $o'.ext \wedge p(o')$ to $R^+$. 

The macros also avoid introducing unnecessary definition variables in lemmas, which would subsequently need to be eliminated by applying the \textbf{ApplyLemma} and the \textbf{Unit} rules ({see Fig.~\ref{fig:folrules}}).  

\begin{example}\label{example:proof}
    Let $A$ and $B$ be two classes of relational objects with a single attribute $t$.  For brevity, we write $\forall a$ and $\forall b$ as the shorthand for $\forall a: A$ and $\forall b: B$, respectively. We want to prove the UNSAT of the \tfol formula 
    \begin{equation*}
    \begin{split}  
        \phi := \exists h_1 (\forall h_2 \cdot h_1.t 
\ge h_2.t \wedge (\exists r_1 \cdot r_1.t = -h_2.t)) \wedge \\  (\forall r_2 \cdot r_2.t > h_1.t) \wedge (r_1.t > 0)
\end{split}
    \end{equation*}
    
\noindent The initial state $s^0$ has an empty domain $\domain = \emptyset$, an empty fact set $F = \emptyset$, and $R^+ = \phi$. The derivation steps for UNSAT are shown in Fig.~\ref{fig:proof}. 
Each step derives new resources (colored in blue) based
on \tfol derivation rules.  
 At step 8, a subset of lemmas in $R^+_8$ are communicated as $\mathbb{T}$ facts to $F$, which serves as a sound over-approximation of $\phi$. 
Step 9 determines that the over-approximation is unsatisfiable, and thus
derives $\bot$ to conclude \textbf{UNSAT}
at step 10.

\begin{figure*}[t]
    \centering
    \begin{tabular}{|l| l |l | l | }
        \toprule
        \# & \makecell{\textbf{Name}} & \makecell{Dependencies} & \makecell{Effect}\\
        \hline 
        0 & \textbf{Input} & \makecell{Input $\phi$} & \makecell{$R^+_0 = \{\phi := \exists h_1 (\forall h_2 \cdot h_1.t 
\ge h_2.t  \wedge (\exists r_1 \cdot r_1.t = -h_2.t)) \wedge $  $ (\forall r_2 \cdot r_2.t > h_1.t) \wedge (h_1.t > 0)\}$} \\
        \hline
        1 & \textbf{ExistentialInst*} & \makecell{$R'_1 = \{\phi\}$} & \makecell{$F_1 = \emptyset$, $\domain_1 = \{\blue{h^o_1}\}$ and $R^+_1 = \{\phi, \blue{\phi_{h^o_1}}\}$ where 
        $\phi_{h^o_1}:= h^o_1.ext \wedge (\psi_1) \wedge (\psi_2) \wedge h^0_1.t > 0$, \\
        $\psi_1 := (\forall h_2 \cdot h^0_1.t 
\ge h_2.t \wedge \ldots)$ and $\psi_2:=\forall r_2. r_2.t > h^o_1.t$}\\
        \hline
        2 & \textbf{RewriteAND*} & \makecell{$R'_2 = \{\phi_{h^o_1}\}$} & 
        \makecell{ $F_2 =\emptyset$, $\domain2 = \{h^o_1\}$ and   $R^+_2 = \{\phi, \phi_{h^o_1}, \blue{h^o_1.ext, \psi_1, \psi_2, h^o_1.t > 0}\}$}  \\ 
        \hline
        3 & \textbf{UniversalInst*} & \makecell{$\domain'_3 = \{h^o_1\}  \wedge$ \\ $ R'_3 = \{\phi_1\}$} & 
        \makecell{ $F_3 =\emptyset$, $\domain_3 = \{h^o_1\}$ and  $R^+_3 = \{\phi, \phi_{h^o_1}, h^o_1.ext, \psi_1, \psi_2, h^o_1.t > 0, \blue{\psi_3}\}$ where \\
        $\psi_3 := h^o_1.ext \Rightarrow \psi_4$ and 
        $\psi_4 := h^o_1.t \ge h^o_1.t \wedge \exists r_1 \cdot r_1.t = h^o_1.t$} \\
        \hline 
        4 & \textbf{Unit} & \makecell{$R'_4  = \{\psi_3, h^0_1.ext\}$}  & \makecell{ $F_4 =\emptyset$, $\domain_4 = \{h^o_1\}$ and $R^+_4 = \{\phi, \phi_{h^o_1}, h^o_1.ext, \psi_1, \psi_2, h^o_1.t > 0, \psi_3, \blue{\psi_4}\}$ } \\
        \hline 
        5 & \textbf{RewriteAND*} & \makecell{$R'_5 =\{\psi_4\}$ } & \makecell{ $F_5 =\emptyset$, $\domain_5 = \{h^o_1\}$ and $R^+_5 = \{\phi, \phi_{h^o_1}, h^o_1.ext, \psi_1, \psi_2, h^o_1.t > 0, \psi_3, \psi_4, \blue{\psi_5}\}$ \\ 
        where $\psi_5 := \exists r_1 \cdot r_1.t = h_1^o.t$} \\
        \hline
        6 & \textbf{ExistentialInst*} & \makecell{$R'_6 = \{\psi_5\}$} & \makecell{ $F_6 =\emptyset$, $\domain_6 = \{h^o_1, \blue{r^o_1} \}$ and  $R^+_6 = \{\phi, \phi_{h^o_1}, h^o_1.ext, \psi_1, \psi_2, h^o_1.t > 0, \psi_3, \psi_4, \psi_5, \blue{\psi_6}\}$ \\ 
        where $\psi_6 := r^o_1.ext \wedge r^o_1.t = -h^o_1.t$} \\ 
        \hline
        7 & \textbf{UniversalInst*} & \makecell{$R'_7 = \{\psi_2\} \wedge$ \\ $\domain'_7 = \{r^o_1\}$} & \makecell{ $F_7 =\emptyset$, $\domain_7= \{h^o_1, r^o_1 \}$ and  $R^+_7 = \{\phi, \phi_{h^o_1}, h^o_1.ext, \psi_1, \psi_2, h^o_1.t > 0, \psi_3, \psi_4, \psi_5, \psi_6,\blue{\psi_7}\}$ \\ 
        where $\psi_7 := r^o_1.ext \Rightarrow r^o_1.t > h^o_1.t$} \\ 
        \hline 
        8 & $\tfolraw \rightarrow \mathbb{T}$ & \makecell{$R'_8 = \{\psi_6,$ \\ $ \psi_7, h^o_1.t >0\}$} &  \makecell{ $F_8 =\{\blue{\psi_6, \psi_7, r^o_1.t > h^o_1.t}\}$, $\domain_8 = \{h^o_1, r^o_1 \}$ and \\ $R^+_8 = \{\phi, \phi_{h^o_1}, h^o_1.ext, \psi_1, \psi_2, h^o_1.t > 0, \psi_3, \psi_4, \psi_5, \psi_6, \psi_7\}$} \\
        \hline
        9 & $\mathbb{T}$-$\textbf{Derive}$ & \makecell{$F'_9 = \{F_8\}$} & \makecell{ $F_9 =\{\psi_6, \psi_7, r^o_1.t > h^o_1.t, \blue{\bot}\}$, $\domain_9 = \{h^o_1, r^o_1 \}$ and \\ $R^+_9 = \{\phi, \phi_{h^o_1}, h^o_1.ext, \psi_1, \psi_2, h^o_1.t > 0, \psi_3, \psi_4, \psi_5, \psi_6, \psi_7\}$ } \\
        \hline
        10 & \textbf{UNSAT} & \makecell{$F'_{10} = \{\bot\}$} & \makecell{\blue{\textbf{UNSAT}}} \\
        \hline
    \end{tabular}
    
    \caption{{\small Proof of UNSAT for $\phi$ in Example~\ref{example:proof}. For each derivation step,  $\#$ is the step number, \textbf{Name} is the name of the applied derivation rule, Dependencies is the dependant resources $Dep'$ for each step, and Effect is the resulting state after the derivation. The resources in blue are newly derived in each step.}}
    \label{fig:proof}
    \vspace{-0.15in}
\end{figure*}

\end{example}

\section{Checking \tfol Proof of  Unsatisfiability}\label{sec:proofchecking}
Let a proof consisting of a sequence of derivation steps $D := d_1 \ldots d_n$, and the initial state $s_0 = (\domain_0, R^+_0, F_0)$ be given, where $R^+_0 = R$ is the set of input formulas, and $\domain_0 = F_0 = \emptyset$ is an empty set of SEOs and facts, respectively. A simple method (denoted as \textit{forward-checking}) to verify the proof is to apply every derivation step $d_i \in D$ from the initial state $s_0$ in order. The proof is \emph{valid} if every derivation step $d_i$ is enabled in state $s_{i-1}$, and the final derivation step $d_n$ applies the rule \textbf{UNSAT}.

The forward checking procedure requires enumerating every derivation step while carrying over the entire state information, which is an expensive procedure.  We observe that: (1) not every derivation step in the proof contributes to the derivation of \textbf{UNSAT}, and (2) only a small portion of the state information is necessary to check a derivation step. These observations enable a potential performance gain by keeping only relevant resources (i.e., lemmas, facts, and SEOs) for the derivation of UNSAT and skipping those steps that do not produce these resources. Inspired by Heule et al.~\cite{DBLP:conf/fmcad/HeuleHW13}, we propose Alg.~\ref{alg:backwardCheck} to check and trim  
an \tfol proof \emph{backwards} from the derivation of \textbf{UNSAT}.

Alg.~\ref{alg:backwardCheck} takes a set of \tfol formulas $R$ and 
 {a  
proof of \textbf{UNSAT} of $R$}, $S = {d_1, \ldots, d_n}$ as input and returns true if and only if $S$ is a valid proof of $R$. If the proof is successfully checked, Alg.~\ref{alg:backwardCheck} computes the subset of derivations that are sufficient to derive \textbf{UNSAT} and stores them {as the proof \textit{core}}. 
Alg.~\ref{alg:backwardCheck} uses \textit{core} to keep track of the derivation steps on which the correctness of the proof depends. Initially, \textit{core} contains only the final derivation of \textbf{UNSAT} $d_n$ (line~\ref{ln:initDn}). Then Alg.~\ref{alg:backwardCheck} proceeds to check the proof backwards from $d_n$ (line~\ref{ln:startDn}). For every derivation step $d_i$, Alg.~\ref{alg:backwardCheck} first checks whether the proof depends on $d_i$ (line~\ref{ln:checkcore}), and skips the derivation step if it does not.

For every dependent step $d_i$, Alg.~\ref{alg:backwardCheck} checks the rule-specific condition on $d_i$ (line~\ref{ln:pccheck}) and returns false if the check fails (line~\ref{ln:pccheckfail}). If the rule-specific condition is successfully checked for $d_i$, then the listed dependent resources $Dep_i$ in $d_i$ are sufficient to prove $d_i$. Alg.~\ref{alg:backwardCheck} then checks whether the dependency $Dep_i$ can be further reduced by invoking the procedure $\textit{Minimize}_{rule_i}$ on $d_i$ and $Dep_i$ (line~\ref{ln:minimizeDP}). Note that the minimization procedure depends on the derivation rule $rule_i$ of $d_i$ which we briefly discuss later in the section.

After minimizing the dependent resources, Alg.~\ref{alg:backwardCheck} finds the derivation steps that derive the dependent resources prior to the application of $d_i$ and adds them to \textit{core} (line~\ref{ln:addcore}). If some dependent resources cannot be derived by any derivation step $d_j$ where $j < i$, and the resource is not given as input (line~\ref{ln:inputcheck}), then Alg.~\ref{alg:backwardCheck} returns \textit{False}. After updating \textit{core}, Alg.~\ref{alg:backwardCheck} proceeds to check the next derivation step in the \textit{core}. The algorithm terminates when every step in \textit{core} has been checked, and at this point, \textit{core} contains a subset of derivation steps which can derive the unsatisfiability of $R$.  

Alg.~\ref{alg:backwardCheck} calls procedure $\textit{Minimize}_{rule_i}$ to minimize the dependent resources $Dep_i$ for checking the validity of a derivation step $d_i$. The minimizing procedures can be trivially derived from the definition of $PC_{rule_i}$ with one exception: the rule $\mathbb{T}$-$\textbf{Derive}$. Recall that $PC_{\mathbb{T}-\textbf{Derive}}$ states that a fact $\phi_g$ can be derived by $d_i$ if $F'_{i} \models^{\mathbb{T}} \phi_g$, where $F'_{i}$ is the set of facts in the dependent resources $Dep_i$ listed by $d_i$. Therefore, $\textit{Minimize}_{\mathbb{T}-\textbf{Derive}}$ can reduce $F'_i$ to a subset $F''_{i}$ if $F''_{i} \models^{\mathbb{T}} \phi_g$.

There are many ways to compute $F''_i$ with trade-offs between performance and minimality (e.g., {by computing} the UNSAT core or the minimum unsatisfiability subset of $F' \wedge \neg \phi_g $) 
. {In this work, we chose to minimize $F'$ by incrementally refine the UNSAT \emph{core} until a fixpoint or a timeout have been reached.}

\begin{algorithm}[t]
                	\caption{Determines whether $S$ is a valid proof of UNSAT for input $R$. If the proof is successfully checked, the computed \emph{core} contains the subset of derivations sufficient to prove UNSAT.
                 }
                	  \small
                	  \hspace*{\algorithmicindent} {\small \textbf{Input:} a set of  \tfol formulas $R$, an \tfol proof $S= d_1 \ldots d_n$.\hfill\mbox{}}\\
                	  \hspace*{\algorithmicindent} {\small \textbf{Output:} $\top$ if and only if $S$ is a valid proof of UNSAT for $R$.\hfill\mbox{}}\\
                   \hspace*{\algorithmicindent} {\small \textbf{Side effect:} \emph{core} stores a subset of $S$ sufficient to prove UNSAT.\hfill\mbox{}
                   }                 
            
            \begin{algorithmic}[1]
                	    \State \textit{core} $\gets \{d_n\}$ \label{ln:initDn}
                            \For {$i \in (n, n-1 \ldots, 1)$}\label{ln:startDn}
                                \If {$d_i \not\in \textit{core}$} {\textbf{Skip}} \EndIf \label{ln:checkcore}
                                \State $(rule_i, R^\Delta_i, F^\Delta_i, \domain^\Delta_i, Dep_i) \gets d_i$
                                \If {$Pc_{rule_i}(d_i)$} \label{ln:pccheck}
                                    \State $Dep* \gets Minimize_{rule_i}(d_i, Dep_i)$
                                    \For{$rec \in Dep*$} \label{ln:minimizeDP}
                                        \If {$rec \in R$} {\textbf{Skip}} \EndIf \label{ln:inputcheck}
                                        \If {$\exists j < i \cdot d_j \text{ derive } rec $}
                                            {$core \uplus d_j$} \label{ln:addcore}
                                         {\Return $\bot$} \label{ln:rescheckfail}
                                        \EndIf
                                    \EndFor
                                \Else
                                \State \Return $\bot$ \label{ln:pccheckfail}
                                \EndIf
                            \EndFor
                        \State \Return $\top$
                	\end{algorithmic} 
                	\label{alg:backwardCheck}
                \end{algorithm}

\begin{example}
     {Continuing from Example~\ref{example:proof}}, suppose we apply Alg.~\ref{alg:backwardCheck} on the input formula $\phi$ and its proof of UNSAT. {Alg.~\ref{alg:backwardCheck} starts from the final derivation step (step 10) which derives UNSAT}: 
     \begin{enumerate}
         \item By checking step 10, the fact $\bot$ is a dependant resource, and step 9 that derived $\bot$ via $\mathbb{T}$-$\textbf{Derive}$ is added to \emph{core}.
         \item {By checking step 9}, the dependant resource $F'$ is reduced to $F_8 = \{\psi_6, \psi_7, h^o_1.t >0\}$, derived by step 8. Therefore, step 8, $\tfolraw \Rightarrow \mathbb{T}$, is added to \emph{core}. 
         \item During the checking {of} step 8, lemmas $\{\psi_7, \psi_6, h^o_1.t >0\}$ are required, derived by steps 7, 6 and 2, respectively. 
          Therefore, steps 7, 6 and 2 are added to \emph{core}.
         \item Step 7 depends on $\psi_2$ and $r^o_1$ which are derived by steps 2 and 6, respectively. Since they are already in  \emph{core}, no updates are needed.
         \item Step 6 depends on $\psi_5$, which is derived by step 5. Therefore, step 5 is added to  \emph{core}.
         \item Step 5 depends on $\psi_4$, which is derived by step 4. Therefore, step 4 is added to  \emph{core}. 
         Step 4 depends on $h^o_1.ext$ and $\phi_3$, which are derived by steps 1 and 3, respectively. Therefore, steps 1 and 3 are added to \emph{core}. 
         \item Step 3 depends on $\psi_1$ and $h_1^0$, which are derived by step 1. Since step 1 is already in \emph{core}, no changes are needed.
         \item Step 2 depends on $\phi_{h^o_1}$ which is derived in step 1. 
 Since step 1 is already in \emph{core}, no changes are needed.
         \item Step 1 depends on $\phi$ which is given as an input. 
        \item All of the derivation steps in \emph{core} have been checked, and the proof is successfully verified.  \emph{Core} contains every step of the original proof. 
     \end{enumerate}
\end{example}
\section{Diagnosing \tfol Unsatisfiability} \label{sec:dignoise}

Suppose that $\phi$ is an unsatisfiable \tfol formula. Understanding the causes of unsatisfiability can be a non-trivial task if $\phi$ is large and complex. Fortunately, the proof provides hints on how $\phi$ is used to derive UNSAT, and we can leverage the proof to generate a simpler (and smaller) unsatisfiable formula $\phi'$ that better explains the causes of unsatisfiability.

\begin{definition}[Proof-Specific UNSAT Diagnosis]\label{def:dig}
Let an  \tfol formula $\phi$ and its proof $S$ of \textit{UNSAT}  be given. The formula $\phi'$ is a \emph{proof-specific \textit{UNSAT} diagnosis of $\phi$ with respect to $S$} if: (1) $\phi'$ is unsatisfiable, and (2) the proof $S$ can be applied to $\phi'$ to derive the unsatisfiability of $\phi'$.
\end{definition}

Intuitively, a proof-specific UNSAT diagnosis of $\phi$ with respect to $S$ is a {formula $\phi'$} 
such that {$\phi \models \phi'$} and $\phi$ and $\phi'$ share the same causes of unsatisfiability  in $S$. If $\phi$ is a conjunction of a set of \tfol formulas $R$, then $\phi'$ can be extracted as an UNSAT core of $R$. We can use Alg.~\ref{alg:backwardCheck} to easily compute an UNSAT core by returning $\textit{core} \cap R$ at the end of proof checking. However, not every formula in the core is used fully for the derivation of UNSAT, and we can further weaken the $\textit{core}$ as long as
the derivation of $S$ is not affected.
More specifically, $\phi'$ is filtered from $\phi$ by $S$ at the level of boolean atoms $\alpha$, which includes Boolean constants ($\top$ and $\bot$), Boolean variables, Boolean attributes of relational objects, $\textit{term}_1 \cong \textit{term}_2$, $\neg \alpha$, and quantifiers on  relational objects (i.e., $\exists o:r$),
\begin{definition}[Active Atoms]\label{def:activeatom}
Let an \tfol formula $\phi$ and a proof of UNSAT $S$ be given. Let $s_n = (R^+_n, F_n, \domain_n)$ be the final state after applying $S$ to the formula input $\phi$. An atom $\alpha$ is \emph{active in $S$} if and only if one of the following conditions holds on $s_n$:
    \begin{enumerate}
        \item every variable and relational object appeared in $\alpha$ is free in $\phi$, and $\alpha \in R^+_n \vee \alpha \in F_n$;
        \item for every reference to a relational object $o$ in $\alpha$, if $o$ is bounded in $\phi$ (i.e., $\forall o$ or $\exists o$), then there exists a relational object $o' \in \domain_n$ such that $\alpha[o \mapsto o']$ is an active atom;
        \item $\neg \alpha$ appears in NNF($\phi$), and $\neg \alpha$ is active.
        \item if $\alpha = \exists o: r \vee \alpha = \forall o: r $, then there exists a relational object $o' \in \domain_{n}$ that instantiates $o$, and $o'.ext$ is active. 
    \end{enumerate}
\end{definition}
Def.~\ref{def:activeatom} defines the criteria for determining which atoms in $\phi$ are used in the derivation of UNSAT. Condition (1) states that an atom without bounded variables is active if and only if it appears either as an \tfol lemma or a $\mathbb{T}$ fact (base case). Condition (2) considers the case when $\alpha$ contains bounded variables introduced by attributes of the bounded relational object $o$. $\alpha$ is active if there exists an instantiation $o' \in \domain_n$ of the bounded relational object $o$ (by \textbf{ExistentialInst} or \textbf{UniversalInst}) such that the instantiated atom $\alpha[o \mapsto o']$ is active. Condition (2) allows us to map atoms in nested quantifiers to the instantiated atoms  used for derivation. Condition (3) covers negation, where $\alpha$ is rewritten as $\neg \alpha$ via \textbf{RewriteNeg}. Condition (4) covers the case of quantifiers: a quantifier is \emph{active} if {the resulting atoms from instantiating the quantifier are active. }
Conjunctions and disjunctions are broken down into atoms (with \textbf{RewriteAND} and \textbf{RewriteOR}, respectively) where the activeness of atoms is checked separately.

\begin{definition}[Atom-level Proof-Specific Diagnosis]\label{def:atomdig}
     Let an UNSAT \tfol formula $\phi$ and its proof of \textit{UNSAT} $S$ be given. The \emph{atom-level proof-specific diagnosis of $\phi$} is the result of substituting every non-active atom in $\phi$ with $\top$.
\end{definition}

\begin{theorem}\label{thm:correctness}
     Let an unsatisfiable \tfol formula $\phi$ and its proof of \textit{UNSAT} $S$ be given. The atom-level proof-specific diagnosis of $\phi$ (Def.~\ref{def:atomdig}) is a proof-specific diagnosis of $\phi$ (Def.~\ref{def:dig}).
\end{theorem}

The proof of Thm.~\ref{thm:correctness} depends on the fact that an atom (or an instantiation of the atom) is used in the derivation of UNSAT only if it appears either as a lemma or a fact in the final state. This is true because the lemma set $R^+$ and the fact set $F$ are monotonically increasing. An atom can be abstracted into $\top$ without affecting the derivation if (1) it has never been recorded as a lemma or a fact, or (2) it has never been used to instantiate lemmas or facts.

\begin{example}\label{example:dig}
    Continuing from  Example~\ref{example:proof},  $\phi$ contains the following atoms: (1) $h_1.t 
\ge h_2.t$, (2) $ r_1.t = -h_2.t$, 
(3) $r_2.t > h_1.t$, (4) $\wedge (h_1.t > 0)$, (5) $\exists h_1$, (6) $\exists r_1$, and (7) $\forall r_2$. Given the proof $S$, atom (2) is active because $\psi_6 \in F_n$; atom (3) is active because $\psi_7 \in F_n$; and atom (4) is active because $h^o_1 > t \in F_n$. Atoms (5), (6) and (7) are active because the relational objects $h^o_1$ and $r^o_1$ are in $\domain_n$ while atoms $h^o_1.ext$ and $r^o_1.ext$ are active.  However, atom (1) is not active, and the atom-level proof-specific diagnosis of $\phi$ is:
    \[\phi_{dig} := \exists h_1 (\forall h_2 (\exists r_1 \cdot r_1.t = -h_2.t)) \wedge  (\forall r_2 \cdot r_2.t > h_1.t) \wedge (h_1.t > 0). \]
\end{example}
In practice, given an \tfol formula $\phi$ and a proof of UNSAT $S$, we first check the proof using Alg.~\ref{alg:backwardCheck} while computing a trimmed 
proof $S'$ extracted from \textit{core}. Then, we compute the atom-level proof-specific diagnosis on $\phi$ and $S'$ to explain a more precise cause of unsatisfiability.

{In case \tfol formulas are compiled from a different source formalism (e.g., SLEEC rules~\cite{feng-et-al-24}), we can  compute the proof-based diagnosis at the level of the source formalism from the \tfol diagnosis by tracking the relationship between atoms in the source formalism and \tfol and projecting the \tfol diagnosis onto the source formalism. In Sec.~\ref{sec:evaluation}, we show this process for SLEEC rules.}
%
\begin{algorithm}[t]
                	\caption{\small $G$: Ground a NNF \tfol  formula $\groundinput$ in a domain $\domainUnder$ \blue{while emitting the \tfol proof}. }
                	  \small
                	  \hspace*{\algorithmicindent} {\small \textbf{Input:} an \tfol formula $\groundinput$ in NNF, and a domain of relational objects $\domainUnder$. \hfill\mbox{}}\\
                	  \hspace*{\algorithmicindent} {\small \textbf{Output:} a grounded quantifier-free formula  $\phi_g$ over relational objects.\hfill\mbox{}}\\
                	\begin{algorithmic}[1]
                	    
                	    \If {match ($\groundinput$, $\exists o : r \cdot \phi'_f)$} \label{ln:extdiscovery} {\color{red}\textit{//process the existential operator}}
                	        \State $o' \gets \textsc{NewAct}(r)$  {\color{red}\textit{//create a new relational object of class $r$}}\label{ln:extnewClass}
                	        \State $\phi_{o'} \gets$ $o'.ext \wedge \phi' [o \gets o']$) \label{ln:extreplacement}
                            \State \blue{\textbf{ExistentialIns*} $R^\Delta = \{\phi_{o'}\}$, $D^\Delta = \{o'\}$,  $Deps = \{R'=\phi\}$}
                            \State $\phi_g \gets \groundAlg(\phi_{o'}, \domain')$
                	    \EndIf
                	    \If {match ($\groundinput$, $\forall o : r \cdot \phi'_f)$} \label{ln:unidiscovery} {\color{red}\textit{//process the universal operator }}
                        \For{$o_i':r \in \domainUnder$} 
                        \State \blue{$\phi_{o_i'} \gets $ $o'.ext\Rightarrow \phi'_f$[$o \gets o_i'$]}
                        \State \blue{\textbf{UniverisalIns*} $R^\Delta = \{\phi_{o_i'}\}$, $D^\Delta = \{o'\}$,  $Deps = \{R'=\phi\}$}
                        \EndFor
                	        \State  \begin{varwidth}[t]{\linewidth}
                	                    $\phi_g \gets \bigwedge_{o_i':r \in \domainUnder}\groundAlg(\phi_{o_i}, \domainUnder)$ \label{ln:forall}
                	                    \end{varwidth}
                	    \EndIf
                        \If {match ($\groundinput$, $\phi_1 \wedge \phi_2$)} {\color{red}\textit{//process the AND operator }}
                        \State \blue{\textbf{RewriteAND*} $R^\Delta = \{\phi_1, \phi_2\}$, $Deps = \{R' = \phi\}$}
                        \State $\phi_g \gets \groundAlg(\phi_1, \domainUnder) \wedge \groundAlg(\phi_2, \domainUnder)$
                        \EndIf
                     \If {match ($\groundinput$, $\phi_1 \vee \phi_2$)} {\color{red}\textit{//process the OR operator }}
                        \State \blue{\textbf{RewriteOr*} $R^\Delta = \{DL^+(\phi_1), DL^+(\phi_2),\defi{\phi_1}\vee \defi{\phi_2}\}$, $Deps = \{R' = \phi\}$}
                        \State $\phi_g \gets \groundAlg(\phi_1, \domainUnder) \vee \groundAlg(\phi_2, \domainUnder)$
                        \EndIf
                        \State \Return $\phi_g$ 
                	\end{algorithmic} 
                	\label{alg:ground}
                	\vspace{-0.05in}
                \end{algorithm}

\section{Supporting Proof of UNSAT for \ibs}
\label{sec:LEGOsSupport}
Here we introduce our extension to the \tfol satisfiability checking algorithm \ibs~\cite{feng-et-al-23} to generate the \tfol proof of UNSAT.

Recall from Sec.~\ref{sec:background} that an \tfol formula is satisfiable if there is some domain $\domain$ and a valuation function $v$ such that $(\domain, v)\models \phi$.
\ibs can detect the unsatisfiability of a given formula $\phi$ if an over-approximation $\phi_g := \groundAlg(\phi, \domain')$ is unsatisfiable for any domain
$\domain'$, where $G$ is defined in Alg.~\ref{alg:ground}. The over-approximation $\phi_g$ is a QFLIA formula whose satisfiability can be decided by an SMT solver. To extend \ibs to support the proof of UNSAT, we need to show: (1)  the derivation steps to derive $\phi_g$, and (2) the  unsatisfiability of $\phi_g$. The latter is trivial as we can 
easily apply the rule $\mathbb{T}$-\textbf{Derive} on $\phi_g$ to derive $\bot$ if $\phi_g$ is UNSAT.  To show the derivation of $\phi_g$, we extend $\groundAlg$  (Alg.~\ref{alg:ground})  to capture the encoding steps as \tfol derivations. The extensions are colored in blue.

\emph{$\groundAlg$} (Alg.~\ref{alg:ground}) requires that the input \tfol formula $\phi$ is in negation normal form (NNF). To satisfy the requirement, we  recursively apply \textbf{RewriteNeg} (see Sec.~\ref{sec:proofFramework}) to convert $\phi$ into NNF. $\groundAlg$ then matches the top-level logical operator ($\exists$, $\forall$, $\wedge$, or $\vee$) of $\phi$ to encode  $\phi_g$ accordingly. To support proof generation, we attach applications of different proof rules (shown in blue) for different cases. For example, when $\groundAlg$ is given $\phi$ that matches the expression $\exists o:r \cdot \phi'$, then $\groundAlg$ first instantiates $o$ with a fresh relational object of class $r$ and then encodes an intermediate formula $\phi_{o'}$ as $o'.\textit{ext} \wedge \phi' \gets [o \leftarrow o']$. To prove that $\phi_{o'}$ is derived from $\phi$, the rule \textbf{ExistentialIns*} is applied on lemma $\phi$ to derive $\phi_{o'}$ and a new relational object $o'$. In case $\phi_{o'}$ contains other quantifiers, $\groundAlg$ is recursively called on $\phi_{o'}$, which would invoke more \tfol derivation rules.

%
\begin{table}[t]
    \centering
    \caption{\small{{Efficiency of \approach in generating UNSAT proofs of \emph{vacuous conflicts} \& effectiveness in trimming them.  t-overhead is the time overhead for generating and checking the proof {and t-check is the time for checking the generated proof}; trimmed/initial is the size of the trimmed and initial  proof files measured in Kilobyte (size) and as the percentage of reduction (\%). n-steps is
    the trimmed and the initial number of proof derivation steps, with \% as the percentage of reduction after checking the proof.}}}
    %
    \label{tab:vc}
    \scalebox{0.8}{
        \begin{tabular}{r c c c c}
        \toprule
             case studies  & t-overhead & t-check & size & n-steps \\
                         & & & trimmed/initial (\%)& trimmed/initial (\%)\\
        \toprule
        Tabiat  & -0.07 & 0.08 &  18/230 (92\%)& 93/963 (90\%)\\ 
        & 0.05 & 0.07 & 18/133 (86\%)& 73/398 (82\%)\\ 
        & -0.04 & 0.1 & 22/256 (91\%)& 114/1091 (90\%)\\ 
        & 0.03 & 0.08 & 18/233 (91\%)& 93/963 (90\%)\\ 
        & 0.01 & 0.08 & 18/245 (92\%)& 93/1028 (91\%)\\ 
        & -0.016 & 0.08 & 18/234 (92\%)& 93/963 (90\%)\\ 
        \bottomrule
        \end{tabular}
    }
\end{table}

\begin{table}[t]
    \centering
    \caption{\small{Efficiency of \approach in generating UNSAT proofs of \emph{situational conflicts} \& effectiveness in trimming them. Significant overhead is in bold.}}
    \label{tab:sc}
    \scalebox{0.8}{
        \begin{tabular}{r c c c c}
        \toprule
             case studies  & t-overhead & t-check &  size & n-steps \\
                         & & & trimmed/initial (\%)& trimmed/initial (\%)\\
        \toprule
        ALMI & 0.13 & 0.09 & 104/256 (59\%)& 137/664 (79\%)\\ \toprule
        ASPEN  & 0.26 & 0.18 & 76/392 (81\%)& 118/1607 (93\%)\\ 
        & 0.21 & 0.16 & 76/230 (67\%)& 111/745 (85\%) \\
        & -0.05 & 0.17 & 76/268 (72\%)& 111/832 (87\%) \\
        \toprule
        DAISY  & 0.33 & 0.23 & 118/648 (82\%)& 122/1854 (93\%) \\ 
        & 0.29 & 0.21 & 117/520 (78\%)& 122/1470 (82\%) \\
        \toprule
        SafeSCAD  & 0.25 & 0.21 & 74/234 (68\%)& 116/648 (82\%) \\ 
        & 0.22 & 0.18 & 68/240 (72\%)& 67/779 (91\%) \\
        & 0.23 & 0.19 & 67/226 (70\%)& 81/721 (89\%) \\
        & \textbf{2.25} & 2.17 & 88/244 (64\%)& 162/843 (81\%) \\
        & 0.34 & 0.27 & 100/358 (72\%)& 204/1175 (83\%) \\
        & 0.26 & 0.24 & 83/196 (58\%)& 156/544 (71\%) \\
        & 0.29 & 0.24 & 88/236 (63\%)& 162/793 (80\%) \\
        & 0.27 & 0.24 & 89/197 (55\%)& 162/544 (70\%) \\
        & 0.32 & 0.29 & 101/248 (59\%)& 207/674 (69\%) \\
        & 0.56 & 0.24 & 83/245 (66\%)& 156/849 (82\%) \\
        & 0.33 & 0.28 & 101/263 (62\%)& 207/769 (33\%) \\
        \toprule
        Tabiat  & 0.13 & 0.09 & 79/203 (61\%)& 55/555 (90\%)\\ 
        & 0.13 & 0.09 & 71/232 (69\%)& 34/586 (94\%)\\ 
        & 0.13 & 0.09 & 68/229 (70\%)& 42/632 (93\%)\\ 
        & 0.14 & 0.10 & 90/199 (55\%)& 105/542 (81\%)\\ 
        & 0.13 & 0.10 & 91/200  (54\%)& 105/542 (81\%)\\ 
        & 0.14 & 0.10 & 84/199 (58\%)& 98/542 (82\%)\\ 
        & 0.14 & 0.10 & 85/244 (65\%)& 93/657 (86\%)\\ 
        & 0.13 & 0.08 & 69/257 (73\%)& 34/699 (95\%)\\ 
        & -0.19 & 0.14 & 98/369 (73\%)& 127/990 (87\%)\\ 
        \bottomrule
        \end{tabular}
    }
\end{table}

\begin{table}[t]
    \centering
    \caption{\small{{Efficiency of \approach in generating UNSAT proofs of \emph{redundancy} \& effectiveness in trimming them.}}}
    \label{tab:red}
    \scalebox{0.8}{
        \begin{tabular}{r c c c c}
        \toprule
             case studies  & t-overhead & t-check &  size & n-steps \\
                         & & & trimmed/initial (\%)& trimmed/initial (\%)\\
        \toprule
        ASPEN  & 0.08 & 0.08 & 27/151 (82\%)& 101/751 (87\%)\\ \toprule
        AutoCar  & 0.21 & 0.15 & 29/143 (80\%)& 97/576 (83\%)\\ 
        & 0.21 & 0.14 & 29/159 (82\%)& 97/673 (86\%)\\ \toprule
        Casper  & 0.17 & 0.14 & 23/85 (73\%)& 93/320 (71\%)\\ \toprule
        CSI-Cobot  & 0.12 & 0.12 & 9/34 (74\%)& 51/131 (61\%)\\ 
        & 0.12 & 0.11 & 9/35 (74\%)& 51/131 (61\%)\\ 
        & 0.13 & 0.07 & 10/52 (81\%)& 53/228 (77\%)\\ \toprule
        DAISY  & 0.15 & 0.14 & 7/48 (85\%)& 30/136 (78\%) \\ \toprule
        SafeSCAD  & 0.02 & 0.01 & 9/39 (77\%)& 51/155 (67\%) \\ 
        & 0.02 & 0.01 & 10/40 (75\%)& 53/159 (67\%)\\ \toprule
        Tabiat  & 0.07 & 0.09 & 19/216 (91\%)& 93/895 (80\%)\\ 
        & 0.09 & 0.07 & 9/50 (82\%)& 44/165 (73\%)\\
        & 0.13 & 0.08 & 15/259 (94\%)& 73/1089 (93\%)\\
        & 0.12 & 0.08 & 19/231 (92\%)& 93/945 (80\%)\\
        & 0.09 & 0.08 & 19/233 (92\%)& 93/947 (80\%)\\
        \bottomrule
        \end{tabular}
    }
\end{table}

\begin{table}[t]
    \centering
    \caption{\small{Efficiency of \approach in generating UNSAT proofs of \emph{restrictiveness} \& effectiveness in trimming them.}}
    \label{tab:res}
    \scalebox{0.8}{
        \begin{tabular}{r c c c c}
        \toprule
             case studies  & t-overhead & t-check & size & n-steps \\
                         & & & trimmed/initial (\%)& trimmed/initial (\%)\\
        \toprule
        SafeSCAD  & 0.04 & 0.02 &  10/43 (77\%)& 53/162 (67\%) \\   
        & 0.03 & 0.01 & 10/42 (77\%)& 60/169 (67\%)\\ \toprule
        Tabiat  & 0.09 & 0.08 & 21/141 (85\%)& 85/455 (81\%)\\  
        & 0.01 & 0.08 & 18/236 (92\%)& 93/1022 (91\%)\\   
        & 0.11 & 0.08 & 18/237 (92\%)& 93/1022 (91\%)\\ 
        \bottomrule
        \end{tabular}
    }
\end{table}

\begin{table*}[t]
    \centering
    \caption{\small{{Usability of the well-formedness diagnosis computed using \approach. For each generated diagnosis of a property violation, the table records the number of rules present in the diagnosis (nRu) and the total number of rules in the case study (nR), the number of clauses highlighted in the rules (nC) and the number of clauses used (nCu) to resolve the issue, the number of events/measures present in the diagnosis (nE/nM) and the number of these events/measures used to resolve the issue (nEu/nMu), and finally, the number of events present  (nSe) and used (nSEu) in the situation to resolve the situational conflicts. }}}
    \label{tab:WFI}
    \scalebox{0.8}{
        \begin{tabular}{rc cc cccc cc ccc}
        \toprule
             \multirow{3}{*}{case studies} && \multicolumn{11}{c}{well-formedness issues}  \\
             & & \multicolumn{2}{c}{conflict} & \multicolumn{4}{c}{s-conflicts} & \multicolumn{2}{c}{redundancy} & \multicolumn{2}{c}{restrictiveness}\\ 
             && nRu (Ru) & nCu (nC) & nRu (nR) & nCu (nC) & nSEu (nSE) & nSMu (nSM) & nRu (nR) & nCu (nC) & nC (nCu) & nEu (nE) & nMu (nM)\\
             \toprule
             ALMI && \multicolumn{2}{c}{NA} & 2 (39) & 4 (7) & 2 (3) & 1 (2) & \multicolumn{2}{c}{NA} & \multicolumn{3}{c}{NA}\\
             ASPEN && \multicolumn{2}{c}{NA} &  2 (15) & 6 (6) & 2 (3) & 2 (3) &  3 (15) & 7 (7) & \multicolumn{3}{c}{NA}\\
             Casper && \multicolumn{2}{c}{NA} & \multicolumn{4}{c}{NA} & 2 (26) & 8 (8) & \multicolumn{3}{c}{NA}\\
             DAISY && \multicolumn{2}{c}{NA} & 2 (26)  & 6 (6) & 2 (5) & 5 (5) & 1 (1) & 2 (2) & \multicolumn{3}{c}{NA}\\
             CSI-Cobot && \multicolumn{2}{c}{NA} & \multicolumn{4}{c}{NA} & 2 (20) & 4 (4) & \multicolumn{3}{c}{NA}\\
             SafeSCAD && \multicolumn{2}{c}{NA} & 2 (28) & 4 (6) & 1 (5) & 2 (3) & 2 (2) & 4 (4) & 4 (4) & 2 (2) & 1 (1)\\
             Tabiat && 4 (28) & 1 (3) & 2 (28) & 5 (6) & 2 (4) & 3 (3) & 4 (28) & 1 (4) & 2 (3) & 2 (4) & 0 (0)\\
            \bottomrule
        \end{tabular}
    }
\end{table*}

\section{Evaluation}
\label{sec:evaluation}


In this section, we first evaluate the efficiency of \approach, the implementation of our approach (see Sec.~\ref{sec:dignoise}) in generating proofs of unsatisfiability, its effectiveness in trimming the proof, and the utility of its \tfol diagnosis.
Specifically, we aim to answer 
\textbf{RQ1:} 
{How efficient is the proof framework for (a) generating, and (b) checking proofs, in terms of the time overhead?}
We utilize \approach to generate proofs, verify them, and measure the overhead associated with both the proof generation and the proof verification (employing a backward-checking mode).
Then, we aim to answer \textbf{RQ2:} How effective is \approach for trimming the proof? 
We  compare the initial size of the proof file and the number of derivation steps, with the ones trimmed.
Finally, we aim to answer \textbf{RQ3:} How well does the instantiated diagnosis assist users in debugging identified inconsistencies? Through answering this question, we aim to determine whether the instantiated diagnosis provides meaningful insights for debugging inconsistencies. This demonstrates the utility of the \tfol diagnosis on a use case. 
\approach implementation and the evaluation artifacts are available in \cite{ASE-Artifact-new}.

\noindent
{\underline{Remark.} In this section, we do not compare our work with existing approaches because no approaches  support the proof of unsatisfiability for \tfol, and a direct comparison with proof support for other formalisms is not feasible, as different formalisms have proof systems and rules with incomparable expressive power.}

\boldparagraph{Case studies}
We considered the \tfol requirements developed for eight real-world case-studies
taken from the RESERVE repository of normative requirements, with \tfol satisfiability used o check their well-formedness~\cite{feng-et-al-24,feng-et-al-24b}: 
(1) ALMI~\cite{Hamilton-et-al-22} -- a system assisting elderly or disabled users in a monitoring/advisory role and with everyday tasks; 
(2) ASPEN~\cite{aspen-23} --  an autonomous agent dedicated to forest protection, providing both diagnosis and treatment of various tree diseases; 
(3) AutoCar~\cite{autocar-20} -- a system that implements emergency-vehicle priority awareness for autonomous vehicles; 
(4) CSI-Cobot~\cite{Stefanakos-et-al-22} -- a system ensuring the safe integration of industrial collaborative robot manipulators; 
(5) DAISY~\cite{daisy-22} -- a sociotechnical AI-supported system that directs patients through an A\&E triage pathway (our running example); 
(6) SafeSCAD~\cite{calinescu2021maintaining} -- a driver attentiveness management system to support safe shared control of autonomous vehicles;
(7) Tabiat~\cite{tabiat} -- a smartphone application (and sensors) that records symptoms and keeps doctors updated on patients' chronic obstructive pulmonary disease conditions; and (8) Casper~\cite{Moro-et-al-18} -- a socially assistive robot designed to help people living with dementia by supporting activities of daily living at home and improving their quality of life.

\begin{figure}
    \centering
    \shadowbox{\includegraphics[scale=0.37]{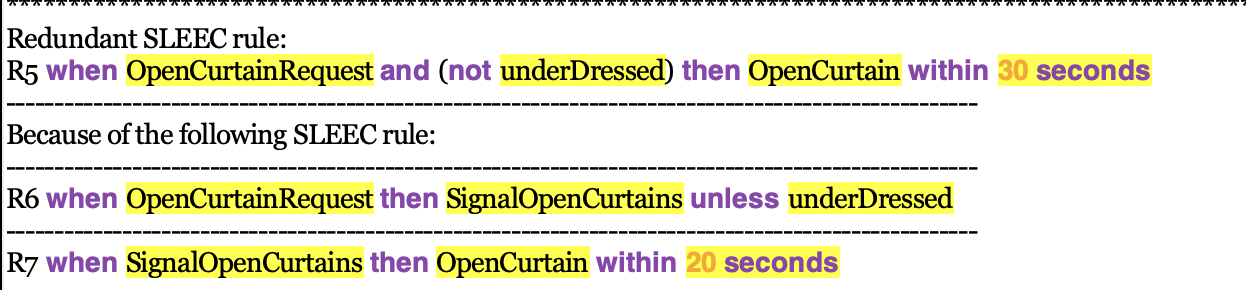}}
    \caption{Redundancy diagnosis example.}
    \label{fig:diagRed}
\end{figure}

\boldparagraph{Well-formedness properties}
\tfol has been used as an underlying formal language for SLEEC, a rule-based language for expressing normative requirements~\cite{Getir-Yaman-et-al-23}. {SLEEC rules  define metric temporal constraints over actions, referred to as \emph{events}, and (unbounded) environmental variables, referred to as \emph{measures}.}  {Here is an example SLEEC rule: \textbf{when} MeetingUser \textbf{and not} patientStressed \textbf{then} ExamineState. Here `MeetingUSer' and `ExamineState' are events, and patientStressed is a measure.}  Given the declarative nature of requirements specification with SLEEC rules, they are prone to having inconsistencies and other
well-formedness properties. 
The following well-formedness properties of SLEEC rules can be verified via \tfol unsatisfiability checking~\cite{feng-et-al-24}: \emph{vacuous conflict}, \emph{situational conflict}, \emph{redundancy}, and \emph{over-restrictiveness}. (i) a rule is  \emph{vacuously conflicting in a rule set $R$} if triggering this rule always leads to a conflict with some other rules in $R$; (ii) a rule is \emph{situationally conflicting in a rule set $R$} if triggering this rule under a certain situation, which defines a history of events and measures, always leads to a conflict with some other rules in the future; (iii) a rule is  \emph{redundant in a rule set $R$} if this rule is logically implied by other rules in $R$; and (iv) a rule set $R$ is \emph{overly restrictive subject to a given purpose} if none of the functional goals represented by the purpose is realizable while respecting $R$.
 Refer to \cite{feng-et-al-24} for formal definitions.

\boldparagraph{RQ1}
To answer RQ1, for each case study and well-formedness property, we measure the overhead (time) of generating and checking the UNSAT proof for the existence of the well-formedness issues. {The results for vacuous conflicts, situational conflicts, redundancy, and restrictiveness are shown in the second column of Tbl.~\ref{tab:vc}, Tbl.~\ref{tab:sc}, Tbl.~\ref{tab:red}, and Tbl.~\ref{tab:res}, respectively.} 
{We found 52 unsatisfiable instances across the eight case studies. 
The average overhead for proof generation was} 0.03 {seconds (with the geometric mean} of 34\%). On the other hand, the geometric mean for proof checking (compared to the solving time without proof generation) was  153\%.  
{Overall},  the overhead for {proof generation} was sufficiently small, enabling the online (i.e., during SAT solving) usage of this technique.
Therefore, the results  answer RQ1.

\boldparagraph{RQ2}
To answer RQ2, for each case study and well-formedness property, we measure the initial proof  size with the number of proof derivation steps, as well as the proof size and the number of steps remaining once trimmed (after checking the proof). {The results for vacuous conflicts, situational conflicts, redundancy, and restrictiveness are shown in the third and fourth columns of Tbl.~\ref{tab:vc}, Tbl.~\ref{tab:sc}, Tbl.~\ref{tab:red}, and Tbl.~\ref{tab:res}, respectively.
The size of the trimmed proof after checking the UNSAT proof generated by \approach is reduced by between 54\% (UNSAT proof for situational conflicts) and 94\% (UNSAT proof for redundancy). The number of steps remaining after checking the proof is reduced by between 33\% (UNSAT proof for situational conflicts) and 93\% (UNSAT proof forredundancy). Overall, \approach is considerably effective for trimming the UNSAT proof, achieving at least a 50\%  reduction in size and a one-third reduction in the number of steps, thereby answering RQ2.}

\boldparagraph{RQ3}
To answer RQ3, we conducted an experiment with five non-technical stakeholders (N-TSs) to determine whether the diagnoses obtained for eight RESERVE real-world case-studies aid in understanding and resolving the specification inconsistencies.
The N-TSs included an ethicist, a lawyer, a philosopher, a psychologist, and a medical doctor.   
{We met with the N-TSs in a single session for each case study. They had access to the full set of requirements, the set of violations, and the diagnoses that were produced the different well-formedness issues identified. The N-TSs used each diagnosis to understand the issue and then resolved it by editing, merging, removing, or adding requirements. We recorded which information from the diagnosis was used in making these decisions.} 
Different diagnoses are generated for different well-formedness properties. \nf{For instance, for the assistive robot ALMI, an example of diagnosis generated for redundancy is shown in Fig.~\ref{fig:diagRed}}. Therefore, for each property type, we measured how different  information provided was used by the N-TSs to resolve the well-formedness problems: (a) for vacuous conflict, redundancy, and restrictiveness: the rules, events, and measures identified; and (b) for situational conflict: the rules and the events, and measures in both the rules and the situation. The results are shown in Tbl.~\ref{tab:WFI}. 

The rule sets for the different case studies ranged between 15 and 19 rules.  For each well-formedness issue, the diagnoses identified a small core of 2-4 rules responsible for the violation  
which allowed the stakeholders to focus on these rules and easily resolve the conflicts.
Furthermore, the highlighting of the clauses in the diagnoses turned out to be very useful:  out of the 14 instances, all highlighted clauses were used in the resolution
in 8 cases, and at least one of them in 2 cases. 
\nf{We observed that the highlighted clauses helped the stakeholders understand exactly what caused the issues.   For example, unnecessary redundant rules, such as R5 in the diagnosis in Fig.~\ref{fig:diagRed},  have all of their clauses highlighted and are present (and highlighted) in other rules within the diagnosis (e.g., rules R6 and R7 in Fig.~\ref{fig:diagRed}).}
At least one of the events/measures present in both the situation and the rules was directly used in the resolution (i.e., present in the added, removed, or updated rules). 
In conclusion, for each type of well-formedness checks, the stakeholders were satisfied with the information presented to them in the diagnosis, agreeing that it provides meaningful insights for helping debug rule inconsistencies, answering RQ3.   The stakeholders suggested going further and recommending resolution patches. Thus, using our UNSAT proof-based diagnosis to compute suggestable resolution patches is a possible future research direction.

\boldparagraph{Summary}
Our experiments demonstrated that \approach is efficient in terms of additional time spent for generating and checking proofs, potentially allowing us to use it in an online manner, during satisfiability checking. 
\approach was also shown to be effective at trimming the proof size (by at least 50\%) and the number of proof steps (by at least a third) while remaining highly useful. Indeed, in our experiments these trimmed proofs were used to instantiate diagnoses, which were shown to stakeholders to resolve identified issues.  Such diagnoses provided them with meaningful insights for debugging inconsistencies between their rules. This demonstrates practical  utility of the \tfol diagnosis.

\boldparagraph{Threats to validity}
{(1) While we only considered eight case-studies, we mitigated this threat by choosing them from different domains. (2) The effectiveness of the generic proof objects derived from FOL* depends on the application. In this study, we only evaluated one application, well-formedness diagnosis for normative requirements. We leave evaluating the effectiveness across a wider range of applications as future work.}

%
\section{Related Work}
\label{sec:relatedwork}
\boldparagraph{Certifying unsatisfiability} Certifying the UNSAT result provided by constraint solvers has been a long standing challenge. UNSAT proofs have been extensively studied in the context of propositional logic based on the  of notion of clause redundancy~\cite{10.1007/978-3-642-39634-2_18,10.1007/978-3-662-54577-5_7,DBLP:conf/cade/Cruz-FilipeHHKS17,DBLP:journals/jar/HeuleKB20,DBLP:conf/sat/BussT19,DBLP:conf/sat/Rebola-Pardo23} (i.e., adding new asserting clauses does not affect the input's satisfiability). Our proof support for \tfol is also based on clause redundancy, but with proof rules designed for \tfol-specific theory reasoning. Similarly, SMT solvers need additional
mechanisms to handle theory reasoning~\cite{barrett2015proofs}.
For example, Z3~\cite{DBLP:conf/tacas/MouraB08} outputs
natural-deduction-style proofs~\cite{DBLP:conf/lpar/MouraB08},
which can be reconstructed inside the interactive theorem prover
Isabelle/HOL~\cite{bohme2009proof,DBLP:conf/itp/BohmeW10}.  Similarly,
veriT~\cite{DBLP:conf/cade/BoutonODF09} produces resolution proof
traces with theory lemmas, and supports
proof reconstruction in both Coq~\cite{DBLP:conf/cpp/ArmandFGKTW11} and
Isabelle~\cite{DBLP:journals/corr/abs-1908-09480,DBLP:conf/cade/BarbosaBF17,barbosa2019better}.
As a more general approach, CVC5~\cite{DBLP:conf/cav/BarrettCDHJKRT11}
produces proofs in the LFSC
format~\cite{DBLP:journals/fmsd/StumpORHT13}, which is a meta-logic that
allows describing theory-specific proof rules for different SMT theories. Similarly to the proof support in SMT solvers, we provided theory-specific proof rules to cover the reasoning steps for soundly deriving the \tfol over-approximation. In contrast, the existing work primarily focused on the quantifier-free fragment of first-order logic and is complementary to our work, as it can be used to close the gap to ensure the correctness of external reasoning steps in our proof framework.

\boldparagraph{Diagnosing unsatisfiability} Identifying the causes of unsatisfiability has been studied in the context of computing UNSAT Cores. DRAT-trim~\cite{DBLP:conf/fmcad/HeuleHW13} enabled UNSAT core production via backward checking of DRAT proof of unsatisfiability for input constraints in propositional logic. Maric et al.~\cite{DBLP:journals/afp/MaricST18} proposed a method to generate UNSAT Cores while running Simplex for solving constraints in linear real arithmetic. Cimatti et al.~\cite{DBLP:journals/jair/CimattiGS11}  proposed a \textit{lemma-lifting approach} to compute UNSAT core in SMT solvers by lazily lifting theory information into boolean abstractions and then refining boolean UNSAT cores backs to the original constraints. Our approach to diagnosis computation can be viewed as an extension of the lemma-lifting approach. Instead of lifting theory information to the boolean level, we lazily lift theory information into the level of a decidable theory $\mathbb{T}$ where UNSAT core production is supported. We then refine the abstracted UNSAT core in $\mathbb{T}$ back to the original constraints in \tfol. Compared to DRAT-Trim, our approach for proof trimming follows a similar backward checking methodology but at a more fine-grained level due to the use of lazy $\mathbb{T}$ abstractions.

\section{Conclusion}
\label{sec:conclusion}
In this paper, we introduced the first method to support proof of unsatisfiability for \tfol by capturing UNSAT deductions through a sequence of verifiable derivation steps. Additionally, we proposed an efficient technique to validate the correctness of these proofs while eliminating unnecessary derivations. From the refined proof, we can derive a concise proof-based diagnosis to explain the cause of unsatisfiability. Our experiments demonstrated the efficiency and effectiveness of our proof support and proof-based diagnosis in software requirements validation tasks. Looking ahead, we aim to enhance the trustworthiness of \tfol proofs by reconstructing them in proof assistants such as Coq. Furthermore, we plan to extend our support to proofs of UNSAT for aggregation functions (e.g., \textit{Sum}, \textit{Count}, \textit{Average}) as additional constructs in \tfol.

\section*{Acknowledgements}
The authors would like to thank the anonymous reviewers for their feedback and insightful comments. We also thank Professor Ana Cavalcanti and Professor Radu Calinescu for their help with the application of \tfol proofs to generate proof-based normative requirements well-formedness diagnoses; the five non-technical stakeholders for participating in our experiments; and finally NSERC and the Amazon Research Award for funding this work.

%
%
\bibliographystyle{splncs04}
\bibliography{ref}

\begin{thebibliography}{10}
\providecommand{\url}[1]{\texttt{#1}}
\providecommand{\urlprefix}{URL }
\providecommand{\doi}[1]{https://doi.org/#1}

\bibitem{Abad-et-al-24}
Abad, P., Aguirre, N., Bengolea, V.S., Ciolek, D.A., Frias, M.F., Galeotti, J.P., Maibaum, T., Moscato, M.M., Rosner, N., Vissani, I.: {Improving Test Generation under Rich Contracts by Tight Bounds and Incremental {SAT} Solving}. In: Proceedings of the sixth International Conference on Software Testing, Verification and Validation, {ICST} 2013, Luxembourg, Luxembourg. pp. 21--30. {IEEE} Computer Society (2013). \doi{10.1109/ICST.2013.46}

\bibitem{DBLP:conf/cpp/ArmandFGKTW11}
Armand, M., Faure, G., Gr{\'{e}}goire, B., Keller, C., Th{\'{e}}ry, L., Werner, B.: {A Modular Integration of {SAT/SMT} Solvers to Coq through Proof Witnesses}. In: Jouannaud, J., Shao, Z. (eds.) Certified Programs and Proofs - First International Conference, {CPP} 2011, Kenting, Taiwan, December 7-9, 2011. Proceedings. Lecture Notes in Computer Science, vol.~7086, pp. 135--150. Springer (2011). \doi{10.1007/978-3-642-25379-9\_12}, \url{https://doi.org/10.1007/978-3-642-25379-9\_12}

\bibitem{autocar-20}
Bahadır, B.N., Kasap, Z.: {AutoCar Project}, \url{https://acp317315180.wordpress.com/}

\bibitem{barbosa2019better}
Barbosa, H., Blanchette, J., Fleury, M., Fontaine, P., Schurr, H.J.: {Better SMT proofs for easier reconstruction}. In: AITP 2019-4th Conference on Artificial Intelligence and Theorem Proving (2019)

\bibitem{DBLP:conf/cade/BarbosaBF17}
Barbosa, H., Blanchette, J.C., Fontaine, P.: {Scalable Fine-Grained Proofs for Formula Processing}. In: de~Moura, L. (ed.) Automated Deduction - {CADE} 26 - 26th International Conference on Automated Deduction, Gothenburg, Sweden, August 6-11, 2017, Proceedings. Lecture Notes in Computer Science, vol. 10395, pp. 398--412. Springer (2017). \doi{10.1007/978-3-319-63046-5\_25}, \url{https://doi.org/10.1007/978-3-319-63046-5\_25}

\bibitem{barrett2015proofs}
Barrett, C., De~Moura, L., Fontaine, P.: {Proofs in satisfiability modulo theories}. All about proofs, Proofs for all  \textbf{55}(1),  23--44 (2015)

\bibitem{DBLP:conf/cav/BarrettCDHJKRT11}
Barrett, C.W., Conway, C.L., Deters, M., Hadarean, L., Jovanovic, D., King, T., Reynolds, A., Tinelli, C.: {CVC4}. In: Gopalakrishnan, G., Qadeer, S. (eds.) Computer Aided Verification - 23rd International Conference, {CAV} 2011, Snowbird, UT, USA, July 14-20, 2011. Proceedings. Lecture Notes in Computer Science, vol.~6806, pp. 171--177. Springer (2011). \doi{10.1007/978-3-642-22110-1\_14}, \url{https://doi.org/10.1007/978-3-642-22110-1\_14}

\bibitem{bohme2009proof}
B{\"o}hme, S.: {Proof reconstruction for Z3 in Isabelle/HOL}. In: 7th International Workshop on Satisfiability Modulo Theories (SMT’09) (2009)

\bibitem{DBLP:conf/itp/BohmeW10}
B{\"{o}}hme, S., Weber, T.: {Fast LCF-Style Proof Reconstruction for {Z3}}. In: Kaufmann, M., Paulson, L.C. (eds.) Interactive Theorem Proving, First International Conference, {ITP} 2010, Edinburgh, UK, July 11-14, 2010. Proceedings. Lecture Notes in Computer Science, vol.~6172, pp. 179--194. Springer (2010). \doi{10.1007/978-3-642-14052-5\_14}, \url{https://doi.org/10.1007/978-3-642-14052-5\_14}

\bibitem{DBLP:conf/cade/BoutonODF09}
Bouton, T., Oliveira, D.C.B.D., D{\'{e}}harbe, D., Fontaine, P.: {veriT: An Open, Trustable and Efficient SMT-Solver}. In: Schmidt, R.A. (ed.) Automated Deduction - CADE-22, 22nd International Conference on Automated Deduction, Montreal, Canada, August 2-7, 2009. Proceedings. Lecture Notes in Computer Science, vol.~5663, pp. 151--156. Springer (2009). \doi{10.1007/978-3-642-02959-2\_12}, \url{https://doi.org/10.1007/978-3-642-02959-2\_12}

\bibitem{DBLP:conf/sat/BussT19}
Buss, S., Thapen, N.: {DRAT} proofs, propagation redundancy, and extended resolution. In: Janota, M., Lynce, I. (eds.) Theory and Applications of Satisfiability Testing - {SAT} 2019 - 22nd International Conference, {SAT} 2019, Lisbon, Portugal, July 9-12, 2019, Proceedings. Lecture Notes in Computer Science, vol. 11628, pp. 71--89. Springer (2019). \doi{10.1007/978-3-030-24258-9\_5}, \url{https://doi.org/10.1007/978-3-030-24258-9\_5}

\bibitem{calinescu2021maintaining}
Calinescu, R., Alasmari, N., Gleirscher, M.: {Maintaining Driver Attentiveness in Shared-Control Autonomous Driving}. In: 16th International Symposium on Software Engineering for Adaptive and Self-Managing Systems (SEAMS). pp. 90--96. IEEE (2021)

\bibitem{daisy-22}
Calinescu, R., Ashaolu, O.: {Diagnostic AI System for Robot-Assisted A\&E Triage (DAISY) website.}, \url{https://twitter.com/NorwichChloe/status/1679112358843613184?t=ALk7s8wcyHztZyyHJoB5pg&s=19}, \url{https://tas.ac.uk/research-projects-2022-23/daisy/}

\bibitem{DBLP:journals/jair/CimattiGS11}
Cimatti, A., Griggio, A., Sebastiani, R.: Computing small unsatisfiable cores in satisfiability modulo theories. J. Artif. Intell. Res.  \textbf{40},  701--728 (2011). \doi{10.1613/JAIR.3196}, \url{https://doi.org/10.1613/jair.3196}

\bibitem{DBLP:conf/cade/Cruz-FilipeHHKS17}
Cruz{-}Filipe, L., Heule, M.J.H., Jr., W.A.H., Kaufmann, M., Schneider{-}Kamp, P.: {Efficient Certified {RAT} Verification}. In: de~Moura, L. (ed.) Automated Deduction - {CADE} 26 - 26th International Conference on Automated Deduction, Gothenburg, Sweden, August 6-11, 2017, Proceedings. Lecture Notes in Computer Science, vol. 10395, pp. 220--236. Springer (2017). \doi{10.1007/978-3-319-63046-5\_14}, \url{https://doi.org/10.1007/978-3-319-63046-5\_14}

\bibitem{10.1007/978-3-662-54577-5_7}
Cruz-Filipe, L., Marques-Silva, J., Schneider-Kamp, P.: {Efficient Certified Resolution Proof Checking}. In: Legay, A., Margaria, T. (eds.) Tools and Algorithms for the Construction and Analysis of Systems. pp. 118--135. Springer Berlin Heidelberg, Berlin, Heidelberg (2017)

\bibitem{aspen-23}
Dandy, N., Calinescu, R.: {Autonomous Systems for Forest ProtEctioN (ASPEN) website.}, \url{https://tas.ac.uk/research-projects-2023-24/autonomous-systems-for-forest-protection/}

\bibitem{feng-et-al-23-b}
Feng, N., Marsso, L., Getir-Yaman, S., Beverley, T., Calinescu, R., Cavalcanti, A., Chechik, M.: {Towards a Formal Framework for Normative Requirements Elicitation}. In: Proceedings of the 38th International Conference on Automated Software Engineering, ({ASE}'2023), Kirchberg, Luxembourg. IEEE (2023)

\bibitem{feng-et-al-24}
Feng, N., Marsso, L., Getir-Yaman, S., Townsend, B., Baatartogtokh, Y., Ayad, R., de~Mello, V.O., Standen, I., Stefanakos, I., Imrie, C., Rodrigues, G., Cavalcanti, A., Calinescu, R., Chechik, M.: {Analyzing and Debugging Normative Requirements via Satisfiability Checking}. In: Proceedings of the 46th International Conference on Software Engineering, (ICSE 2024), Lisbon, Portugal. ACM (2024)

\bibitem{feng-et-al-23}
Feng, N., Marsso, L., Sabetzadeh, M., Chechik, M.: Early verification of legal compliance via bounded satisfiability checking. In: Proceedings of the 34th international conference on Computer Aided Verification ({CAV}'23), Paris, France. Lecture Notes in Computer Science, Springer (2023)

\bibitem{feng-et-al-24b}
Feng, N., Marsso, L., Yaman, S.G., Standen, I., Baatartogtokh, Y., Ayad, R., de~Mello, V.O., Townsend, B., Bartels, H., Cavalcanti, A., Calinescu, R., Chechik, M.: {Normative Requirements Operationalization with Large Language Models}. In: Proceedings of the 32nd IEEE International Conference on Requirements Engineering, RE'2024, Reyjavik Iceland. ACM (2024)

\bibitem{DBLP:journals/corr/abs-1908-09480}
Fleury, M., Schurr, H.: {Reconstructing {veriT} Proofs in {Isabelle/HOL}}. In: Reis, G., Barbosa, H. (eds.) Proceedings Sixth Workshop on Proof eXchange for Theorem Proving, PxTP 2019, Natal, Brazil, August 26, 2019. {EPTCS}, vol.~301, pp. 36--50 (2019). \doi{10.4204/EPTCS.301.6}, \url{https://doi.org/10.4204/EPTCS.301.6}

\bibitem{Getir-Yaman-et-al-23}
Getir-Yaman, S., Burholt, C., Jones, M., Calinescu, R., Cavalcanti, A.: {Specification and Validation of Normative Rules for Autonomous Agents}. In: Proceedings of the 26th International Conference on Fundamental Approaches to Software Engineering ({FASE}'2023), Paris, France. Lecture Notes in Computer Science, Springer (2023)

\bibitem{Hamilton-et-al-22}
Hamilton, J., Stefanakos, I., Calinescu, R., C{\'{a}}mara, J.: Towards adaptive planning of assistive-care robot tasks. In: Proceedings of th Fourth International Workshop on Formal Methods for Autonomous Systems and Fourth International Workshop on Automated and verifiable Software sYstem DEvelopment, (FMAS/ASYDE@SEFM'2022), Berlin, Germany. {EPTCS}, vol.~371, pp. 175--183 (2022). \doi{10.4204/EPTCS.371.12}, \url{https://www.youtube.com/watch?v=VhfQmJe4IPc}

\bibitem{DBLP:conf/fmcad/HeuleHW13}
Heule, M., Jr., W.A.H., Wetzler, N.: Trimming while checking clausal proofs. In: Formal Methods in Computer-Aided Design, {FMCAD} 2013, Portland, OR, USA, October 20-23, 2013. pp. 181--188. {IEEE} (2013), \url{https://ieeexplore.ieee.org/document/6679408/}

\bibitem{DBLP:journals/jar/HeuleKB20}
Heule, M.J.H., Kiesl, B., Biere, A.: Strong extension-free proof systems. J. Autom. Reason.  \textbf{64}(3),  533--554 (2020). \doi{10.1007/S10817-019-09516-0}, \url{https://doi.org/10.1007/s10817-019-09516-0}

\bibitem{King-76}
King, J.C.: Symbolic execution and program testing. Commun. {ACM}  \textbf{19}(7),  385--394 (1976). \doi{10.1145/360248.360252}

\bibitem{DBLP:conf/ifm/KrishnaPS17}
Krishna, A., Poizat, P., Sala{\"{u}}n, G.: {VBPMN:} automated verification of {BPMN} processes (tool paper). In: Polikarpova, N., Schneider, S.A. (eds.) Integrated Formal Methods - 13th International Conference, {IFM} 2017, Turin, Italy, September 20-22, 2017, Proceedings. Lecture Notes in Computer Science, vol. 10510, pp. 323--331. Springer (2017). \doi{10.1007/978-3-319-66845-1\_21}, \url{https://doi.org/10.1007/978-3-319-66845-1\_21}

\bibitem{krishna2019checking}
Krishna, A., Poizat, P., Sala{\"u}n, G.: Checking business process evolution. Science of Computer Programming  \textbf{170},  1--26 (2019)

\bibitem{Hui-et-al-15}
Liang, J.H.J., Ganesh, V., Czarnecki, K., Raman, V.: {SAT-based analysis of large real-world feature models is easy}. In: Proceedings of the 19th International Conference on Software Product Line, {SPLC} 2015, Nashville, TN, USA. pp. 91--100. {ACM} (2015). \doi{10.1145/2791060.2791070}

\bibitem{tabiat}
Liaqat, D.: {The Tabiat website.}, \url{https://www.tabiat.care/}

\bibitem{DBLP:journals/afp/MaricST18}
Maric, F., Spasic, M., Thiemann, R.: An incremental simplex algorithm with unsatisfiable core generation. Arch. Formal Proofs  \textbf{2018} (2018), \url{https://www.isa-afp.org/entries/Simplex.html}

\bibitem{Matos-et-al24}
de~Matos~Pedro, A., Silva, T., Sequeira, T.F., Louren{\c{c}}o, J., Seco, J.C., Ferreira, C.: {Monitoring of spatio-temporal properties with nonlinear {SAT} solvers}. Int. J. Softw. Tools Technol. Transf.  \textbf{26}(2),  169--188 (2024). \doi{10.1007/S10009-024-00740-7}

\bibitem{Moro-et-al-18}
Moro, C., Nejat, G., Mihailidis, A.: Learning and personalizing socially assistive robot behaviors to aid with activities of daily living. ACM Transactions on Human-Robot Interaction (THRI)  \textbf{7}(2),  1--25 (2018)

\bibitem{DBLP:conf/lpar/MouraB08}
de~Moura, L.M., Bj{\o}rner, N.: {Proofs and Refutations, and {Z3}}. In: Rudnicki, P., Sutcliffe, G., Konev, B., Schmidt, R.A., Schulz, S. (eds.) Proceedings of the {LPAR} 2008 Workshops, Knowledge Exchange: Automated Provers and Proof Assistants, and the 7th International Workshop on the Implementation of Logics, Doha, Qatar, November 22, 2008. {CEUR} Workshop Proceedings, vol.~418. CEUR-WS.org (2008), \url{http://ceur-ws.org/Vol-418/paper10.pdf}

\bibitem{DBLP:conf/tacas/MouraB08}
de~Moura, L.M., Bj{\o}rner, N.: {{Z3:} An Efficient {SMT} Solver}. In: Ramakrishnan, C.R., Rehof, J. (eds.) Tools and Algorithms for the Construction and Analysis of Systems, 14th International Conference, {TACAS} 2008, Held as Part of the Joint European Conferences on Theory and Practice of Software, {ETAPS} 2008, Budapest, Hungary, March 29-April 6, 2008. Proceedings. Lecture Notes in Computer Science, vol.~4963, pp. 337--340. Springer (2008). \doi{10.1007/978-3-540-78800-3\_24}, \url{https://doi.org/10.1007/978-3-540-78800-3\_24}

\bibitem{ASE-Artifact-new}
Nick, F., Lina, M.: Supplementary material for ase submission: Diagnosis via proofs of unsatisfiability for first-order logic with relational objects (2024), \url{https://github.com/NickF0211/LEGOS-Proof-Artifact/}

\bibitem{DBLP:conf/sat/Rebola-Pardo23}
Rebola{-}Pardo, A.: Even shorter proofs without new variables. In: Mahajan, M., Slivovsky, F. (eds.) 26th International Conference on Theory and Applications of Satisfiability Testing, {SAT} 2023, July 4-8, 2023, Alghero, Italy. LIPIcs, vol.~271, pp. 22:1--22:20. Schloss Dagstuhl - Leibniz-Zentrum f{\"{u}}r Informatik (2023). \doi{10.4230/LIPICS.SAT.2023.22}, \url{https://doi.org/10.4230/LIPIcs.SAT.2023.22}

\bibitem{Stefanakos-et-al-22}
Stefanakos, I., Calinescu, R., Douthwaite, J.A., Aitken, J.M., Law, J.: Safety controller synthesis for a mobile manufacturing cobot. In: Proceedings of the 20th International Conference on Software Engineering and Formal Methods ({SEFM}'2022), Berlin, Germany. Lecture Notes in Computer Science, vol. 13550, pp. 271--287. Springer (2022). \doi{10.1007/978-3-031-17108-6\_17}

\bibitem{DBLP:journals/fmsd/StumpORHT13}
Stump, A., Oe, D., Reynolds, A., Hadarean, L., Tinelli, C.: {{SMT} proof checking using a logical framework}. Formal Methods Syst. Des.  \textbf{42}(1),  91--118 (2013). \doi{10.1007/s10703-012-0163-3}, \url{https://doi.org/10.1007/s10703-012-0163-3}

\bibitem{10.1007/978-3-642-39634-2_18}
Wetzler, N., Heule, M.J.H., Hunt, W.A.: {Mechanical Verification of {SAT} Refutations with Extended Resolution}. In: Blazy, S., Paulin-Mohring, C., Pichardie, D. (eds.) Interactive Theorem Proving. pp. 229--244. Springer Berlin Heidelberg, Berlin, Heidelberg (2013)

\bibitem{DBLP:conf/pldi/WintererZS20}
Winterer, D., Zhang, C., Su, Z.: Validating {SMT} solvers via semantic fusion. In: Donaldson, A.F., Torlak, E. (eds.) Proceedings of the 41st {ACM} {SIGPLAN} International Conference on Programming Language Design and Implementation, {PLDI} 2020, London, UK, June 15-20, 2020. pp. 718--730. {ACM} (2020). \doi{10.1145/3385412.3385985}, \url{https://doi.org/10.1145/3385412.3385985}

\end{thebibliography}

\begin{ExtendedVersion}
\clearpage
\appendix
\section{\tfol Derivation Rules}
\label{app:folproofsem}

\boldparagraph{Define} If $\phi$ is an \tfol formula and \defi{$\phi$} is undefined, then applying the \textbf{Define} rule assigns $\defi{\phi}$ to a fresh variable and adds the definition clauses $DL^+(\phi)$ and $DL^-(\phi)$ to $R^+$.
    The derivation rule \textbf{Define} allows the introduction of new \tfol formulas in the proof of UNSAT. For example, \textbf{Define} can introduce $\phi$ as a hypothesis, and by deriving $DEF(\phi)$ as a lemma (via $DL^-(\phi)$), we can then use $\phi$ as a lemma (via $DL^+(\phi)$) in the rest of the proof.

    \boldparagraph {Substitute} Suppose an \tfol formula $\phi$ contains a sub-formula $\psi$~\footnote{We assume that bounded variables in $\psi$ can be renamed without causing name collisions.}, where all free variables in $\psi$ are also free in $\phi$ (i.e., $FV(\phi) \supseteq FV(\psi)$). If both $\defi{\phi}$ and $\defi{\psi}$ are defined, then the following lemmas can be added to $R^+$:
    \[\phi^+_{subs}: \defi{\phi} \Rightarrow \phi[\psi \gets \defi{\psi}]\]
    \[\phi^-_{subs}: \overline{\defi{\phi}} \Rightarrow \neg \phi[\psi \gets \defi{\psi}],\]
    where $\phi[\psi\gets \defi{\psi}]$ is the result of substituting $\psi$ with $\defi{\psi}$ in $\phi$.
Note that the condition $FV(\phi) \supseteq FV(\psi)$ prohibits unsound derivation by matching universally quantified variables in $\phi$ to free variables in $\psi$, which are assumed to be existentially quantified.
 Here, we explain the dependencies of the derived lemmas: (1) $\phi_{\text{sub}}^+$ depends on the positive definition of $\phi$ ($DL^+(\phi)$) and the negative definition of $\psi$ ($DL^-(\psi)$); and (2) $\phi_{\text{sub}}^-$ depends on the negative definition of $\phi$ ($DL^-(\phi)$) and the positive definition of $\psi$ ($DL^+(\psi)$).

    \begin{definition}\label{def:subs}
    Let a derivation step $d_i = (\textit{rule}_i, Dep_i, F_i^\Delta, R_i^\Delta, D_i^\Delta)$ be given where $Dep_i = (R'_i, F'_i, \domain'_i)$. Suppose $\phi$ and $\psi$ are two \tfol formula. The step $d_i$ is a valid application of \textbf{substitute} if:

\begin{align*}
  & PC_{\textbf{substitute}}(d_i)  := F^i = \domain_i = \emptyset \wedge \textit{DPos} \vee \textit{DNeg} \text { where }\\
  & \scalebox{0.9}{$\textit{DPos} := ((R^+_i = \{\defi{\phi} \Rightarrow \phi[\psi \gets \defi{\psi}]\} \wedge \{DL^+(\phi), DL^-(\psi))\} \subseteq  D'_i)$}  \\
  & \scalebox{0.9}{$\textit{DNeg}:= (R^+_i = \{ \overline{\defi{\phi}} \Rightarrow \neg  \phi[\psi \gets \defi{\psi}]\} \wedge \{DL^-(\phi), DL^+(\psi))\} \subseteq  D'_i))$}
\end{align*}
\end{definition}

In Def.~\ref{def:subs}, according to the enabling condition $PC_{\textbf{substitute}}$, an application of \textbf{Substitute} derives  either $\phi^+_{subs}$ (following \textit{DPos}) or $\phi^-_{subs}$ (following \textit{DNeg}) one at a time. However, since deriving $\phi^+_{subs}$ does not blocks the derivation of $\phi^-_{subs}$, we assume both lemmas are derived by a single application if not otherwise stated.

    \boldparagraph{ApplyLemma} If $\phi \in R^+$ is a lemma and the definition variable $\defi{\phi}$ is defined, then applying rule \textbf{ApplyLemma} adds lemma $\phi_{\textbf{apply}} := \defi{\phi}$ to $R^+$.
The formal definition of $PC_{\textbf{ApplyLemma}}$ is provided in the appendices. Here, we explain the dependencies of the derived lemma: (1) $\phi_{\textbf{applyLemma}}$ depends on lemma $\phi$.
Intuitively, lemmas in $R^+$ are asserted to be true, and hence their definition variable should also evaluate to true.

    \begin{definition}\label{rule:applylemma}
    Let a derivation step $d_i = (\textit{rule}_i, Dep_i, F_i^\Delta, R_i^\Delta, D_i^\Delta)$ be given where $Dep_i = (R'_i, F'_i, \domain'_i)$. Suppose $\phi$ is  \tfol formula. The step $d_i$ is a valid application of \textbf{ApplyLemma} if:
        \[PC_{\textbf{ApplyLemma}} := R^\Delta_i = \{\defi{\phi}\} \wedge F^\Delta_i = \domain^\Delta_i =\emptyset \wedge \phi \in R'_i \wedge \phi \in \textit{DEF}\]
    \end{definition}

    \boldparagraph{RewriteOR} If an \tfol formula $\phi$ matches $A \vee B$, where $A$ and $B$ are \tfol formulas, and $\defi{\phi}$, $\defi{A}$, and $\defi{B}$ are defined, then applying rule \textbf{RewriteOR} to $\phi$ adds the following lemmas to $R^+$: \[\phi_{or+} := \defi{\phi} \Rightarrow \defi{A} \vee \defi{B},\]
    \[\phi_{or-}^l := \overline{\defi{\phi}} \Rightarrow \overline{\defi{A}} \text{ and } \phi_{or-}^r := \overline{\defi{\phi}} \Rightarrow \overline{\defi{B}}\]

   The dependencies of the derivations are: (1) the lemma $\phi_{\text{or}+}$ depends on the positive definition clauses $DL^+(\phi)$ and the negative definition clauses $DL^-(A)$ and $DL^-(B)$; (2) the lemma $\phi_{\text{or}-}^l$ (and $\phi_{\text{or}-}^r$) depends on the negative definition lemma $DL^-(\phi)$ and the positive definition lemmas $DL^+(A)$ and $DL^+(B)$.

\boldparagraph{RewriteAND} If an \tfol formula $\phi$ matches $A \wedge B$, where $A$ and $B$ are \tfol formulas, and $\defi{\phi}$, $\defi{A}$, and $\defi{B}$ have been assigned. Then applying rule \textbf{RewriteAND} on $\phi$ adds the following lemmas to $R^+$: 
    \[\phi_{and+}^l := \defi{\phi} \Rightarrow \defi{A} \text{ and } \phi_{and+}^r := \defi{\phi} \Rightarrow \defi{B}\]
    \[\phi_{and-} := \overline{\defi{\phi}} \Rightarrow \overline{\defi{A}} \vee \overline{\defi{B}}.\]

    The dependencies of the derivations are: (1) the lemma $\phi_{\text{and+}}^l$ depends on $DL^+(\phi)$ and $DL^-(A)$; (2) the lemma $\phi_{\text{and+}}^r$ depends on $DL^+(\phi)$ and $DL^-(B)$; and (3) the lemma $\phi_{\text{and}-}$ depends on $DL^-(\phi)$, $DL^+(A)$, and $DL^+(B)$.

    \boldparagraph{ExistentialInst} Suppose an \tfol formula $\phi$ matches the pattern $\exists o:r \cdot p(o)$, where $p$ is a \tfol predicate over the free relational object $o$ and $\defi{\phi}$ is assigned. Then applying rule \textbf{ExistentialInst} on $\phi$ adds:
    \begin{enumerate}
        \item a fresh relational object $o'$ ($o' \not\in\domain$) of class $r$ to domain $\domain$,
        \item a lemma  $\phi^+_{Eo'} :=  \defi{\phi} \Rightarrow (o'.ext \wedge p(o'))$ to $R^+$, and 
        \item a lemma  $\phi^-_{Eo'} :=  \overline{\defi{\phi}} \Rightarrow \neg (o'.ext \wedge p(o'))$ to $R^+$
    
    \end{enumerate}
    The dependencies of the derivations are: (1) the relational object $o'$ and the lemma $\phi_{Eo}'$ depend on the positive definition of $\phi$ ($DL^+(\phi)$); (2) the lemma $\phi^-_{Eo'}$ depends on the negative definition of $\phi$ ($DL^-(\phi)$).

    \boldparagraph{UniversalInst} Suppose an \tfol formula $\phi$ matches the pattern $\forall o:r \cdot p(o)$, where $p$ is a \tfol predicate over the free relational object $o$ and $\defi{\phi}$ is assigned. If there exists a relational object $o'$ of class $r$ in $D$, then applying rule \textbf{UniversalInst} on $\phi$ and $o$ adds the following lemma to $R^+$:
    $\phi_{Uo'} :=  \defi{\phi} \Rightarrow (\neg o'.ext \vee p(o'))$.

     The dependencies of the derivations are: the lemma $\phi_{Uo'}$ depends both on the positive definition of $\phi$ ($DL^+(\phi)$) and on the relational object $o'$

\boldparagraph{RewriteNeg} Suppose an \tfol formula matches the pattern $\neg \psi$, where $\psi$ is also an \tfol formula. If $\defi{\phi}$ is defined, then applying \textbf{RewriteNeg} on $\phi$ adds the following lemmas to $R^+$: (1)
    $\phi_{neg}^+ := \defi{\psi} \Rightarrow \textit{NEG}(\phi)$ and (2)
     $\phi_{neg}^- := \overline{\defi{\phi}} \Rightarrow \psi$,  
    where the function \textit{NEG}  pushes the top-level negation into $\psi$ and is defined in a standard way as shown in Fig.~\ref{fig:folrules}.
    \[
    \textit{NEG}(\phi) := \begin{cases}
 \exists o:r \cdot \neg \psi & \text{ if } \phi \text{ matches } \neg \forall o:r \cdot \psi \\
 \forall o:r \cdot \neg \psi & \text{ if } \phi \text{ matches } \neg \exists o:r \cdot \psi \\
 \neg \psi_1 \vee \neg \psi_2 & \text{ if } \phi \text{ matches } \neg (\psi_1 \wedge \psi_b) \\
  \neg \psi_1 \wedge \neg \psi_2 & \text{ if } \phi \text{ matches } \neg (\psi_1 \vee \psi_b) \\
  \phi & \text{otherwise}
    \end{cases}
    \]
{Note that applying \textbf{RewriteNeg} recursively converts an \tfol formula into its negation normal form (NNF) where negation symbols appear in front of atoms (e.g., terms or literals). We can also express \textbf{RewriteAnd} with \textbf{RewriteOr} and \textbf{RewriteNeg}.}

    \boldparagraph{Unit} If there is a lemma $\phi := l \Rightarrow \psi \in R^+$ and a unit literal lemma $l \in R^+$, then applying the \textbf{unit} rule adds $\psi$ as a lemma to $R^+$.
 The dependencies of the derivations are: the lemma $\psi$ depends on both the unit lemma $l$ and $\phi$.

    \boldparagraph{UNSAT} If $\bot$ has been derived as a fact or lemma in the current state, then the {\bf UNSAT} rule can be applied to signal the end of the proof.

\section{Expansion of Derivation Macros}\label{ap:marcoexpansion}
In this section, we show the  expansion for the macro \textbf{RewriteAND*}.

\boldparagraph{RewriteAND*} If a formula $\phi \in R^+$ matches $\psi_1 \wedge \psi_2  \ldots \wedge \psi_n$, then  $\psi_1$, $\psi_2 \ldots \psi_n$ can be directly added to $R^+$. 

    The derivation shortcut can be expanded to the following sequence of the derivation steps: 
    \begin{enumerate}
        \item \textbf{Define} on $\phi$ and $\psi_1 \ldots \psi_n$.
        \item \textbf{Substitute}  $\psi_1 \ldots \psi_n$ in $\phi$ to derive a new lemma $\phi'$, where $\psi_i$ is replaced by a literal $\defi{\psi_i}$.
        \item \textbf{ApplyLemma} on $\phi$ to derive a unit lemma $\defi{\phi}$.
        \item \textbf{Unit}-propagate $\defi{\phi}$ on $\phi'$ to derive $\phi'' := \defi{\psi_1} \wedge \ldots \defi{\psi_n}$.
        \item \textbf{RewriteAND} on $\phi''$ to derive unit lemmas  $\defi{\psi_1} \ldots \defi{\psi_n}$, and 
        \item \textbf{Unit}-propagate $\defi{\psi_1} \ldots \defi{\psi_n}$ on the positive definition lemma $DL^+(\psi_1) \ldots DL^+(\psi_n)$ to derive $\psi_1 \ldots \psi_n$, respectively.
    \end{enumerate}

\end{ExtendedVersion}

\end{document}